# Valley field mechanics: a local perspective beyond valley flavor

Feng-Wu Chen[1], Zheng-Han Huang[1], and Yu-Shu G. Wu[1, 2]

[1] *Department of Electrical Engineering, National Tsing-Hua University, Hsin-Chu 30013, Taiwan*
[2] *Department of Physics, National Tsing-Hua University, Hsin-Chu 30013, Taiwan*

## ABSTRACT

Valleytronics in 2D hexagonal materials is rooted in the existence of Dirac valley flavor, with valley magnetic moments a major resource offered in the materials, which can couple to electric and magnetic fields giving nontrivial field effects important for valley-based applications. Traditionally, such moments have primarily been studied for homogeneous bulk states, from the global perspective with emphasis on the total value of magnetic moment in a state. Rules rigorously established from the perspective, for example, the condition of structural broken inversion symmetry for nonvanishing valley magnetic moments have been widely applied and long guided relevant experiments and applications. However, as demonstrated in this work, hidden degrees of freedom beyond the global perspective exist in the dimension of local physics, with abundant noteworthy twists manifested in the presence of nontrivial structural inhomogeneity with respect to the global perspective. In order to explore the dimension, a $\vec{r}$ -space, Ginzburg-Landau order parameter type, valley-derived field – valley field is introduced. The field describes the local, probability-weighted inversion symmetry breaking instead of the structural one, and has the interpretation of spatial distribution of *local cell-orbital magnetic moments* suited to inhomogeneous structures. A theoretical framework - *valley field mechanics* comprising valley fields and field equations of variant Schrodinger or Klein-Gordon forms is developed to analytically address the local valley physics. Within the framework, the local linear response of a valley field to space-dependent magnetic and electric fields is discussed. It illustrates the existence of *local valley-Zeeman* and *local valley-orbit-interaction* effects and, thus, opens a path to *local valley control* through such effects. Numerical results of valley fields are presented, in bulks, quantum dots, zigzag and armchair quasi-1D structures, of graphene and transition metal dichalcogenides. A variety of intriguing local phenomena are revealed with characteristics in apparent contradiction to global perspective-based expectations and/or constraints, for example,

– *broken "valley flavor* ↔ *magnetic moment orientation" correspondence*,

– *nonvanishing local magnetic moments in the presence of inversion symmetry*,

– *suppressed or even eliminated valley magnetic moments in the presence of broken inversion symmetry.*

By revoking the expectations and/or constraints, the local physics enables a more flexible valley control, including the relaxation of symmetry and material restrictions admitting *gapless, single-layer graphene*, a material with inversion symmetry and manufacturing methods available for routine production of large, high-quality flakes, into the list to ease critical device fabrications in relevant experiments and applications. Overall, the diverse local valley phenomena revealed suggest the exciting direction of *valley field engineering*, e.g., design and search for quantum structures to tailor local valley physics for applications.

## I. Introduction

Following pioneering studies of quantum Hall effect in graphene layers [1–3], atomically thin 2D hexagonal crystals – gapped graphene [4–7] and transition metal dichalcogenides (TMDCs) [8,9] have soon been recognized to form an important class of topological materials [10,11], with a wide spectrum of novel phenomena present in association with the existence of two degenerate and inequivalent band structure valleys (K and K'). Studies have led to the exciting discovery of valley magnetic moments [10], valley Hall effect [10] with nonlocal resistance [12–15], robust valley topological currents [16–22], robust valley-polarized interface states [16], valley selection rule in optical pumping [9,23–26], spin-valley locking [11], valley-orbit [27,28] and valley-Zeeman interactions [10,28], and so on, and have fueled important device proposals for valleytronic applications − valley filters / valves [19,29,30], qubits [28,31–34], FETs [35] and etc. in versatile structures including graphene [36–43] and TMDC [13,44–49] quantum dots (QDs) / quasi-1D (Q1D) structures.

States of topological materials are generally characterized by nontrivial global quantities, for instance, Chern number, $Z_2$ invariant etc., which are topological numbers of ground state manifolds or energy bands in wave vector ( $\vec{k}$ )-space, depending on the material. States in such materials can be topologically protected from limited disturbances, giving rise to phenomenal effects such as extremely long state coherence vital for the applications of spintronics and topological quantum computing [50–53], as well as robust surface metallic states in 2D [54] and 3D [55,56] and end states in Q1D [57–60] in the presence of topological boundaries. In the case of 2D hexagonal crystals, the valley topology is summarized by a 'valley Chern number', essentially the integral of Berry curvature in $\vec{k}$ -space around a valley for a homogeneous bulk. In crystals with broken inversion symmetry, the number is nonvanishing and shows opposite signs between K and K' [16] alluding to the presence of a nontrivial topology with profound impacts, as reflected in particular by the associated physical manifestation of nonvanishing valley-contrasting magnetic moments. Such moments can interact with external fields [10,27,28] giving valley Hall effect, valley-Zeeman and valley-orbit interactions for electric or magnetic valley control important in valley-based experiments and applications.

Traditionally, in analogy to valley Churn numbers, valley magnetic moments have primarily been studied for homogeneous bulk states, from the global perspective with focus on the total value of magnetic moment in a state. Rules rigorously established within the perspective have become popular beliefs, been widely applied, and long guided the field. A famous example is the condition of broken inversion symmetry for nonvanishing valley magnetic moments, with profound impacts on the selection of materials or structures. Take magnetic moment-external field interaction based applications for another instance. As valley magnetic moments are total values, space dependence is excluded and, thus, only uniform external fields have been envisioned and applied. However, from both scientific and application standpoints, given the scarce account of $\vec{r}$ -dependence in the moment, prospective applications and interesting degrees of freedom in valley physics may inadvertently have escaped attention, and rules such as the foregoing ones while contributing to advances may have limited imaginations in the field. Just as in the case of topological materials where topological boundaries engender distinct physics, e.g., robust topological surface states, nontrivial structural inhomogeneity in 2D materials may in principle qualitatively alter valley physics, create intriguing space dependence in the moment, and bring in new prospects. This plausibility motivates us to revisit valley magnetic moment physics and study its space dependence, especially that in confined systems such as Q1D structures and QDs. As the compact topology in confined structures is a strong contrast to the open topology in 2D bulks, fascinating valley phenomena not dictated by valley Chern numbers or valley magnetic moments of bulks may be manifested in confined structures.

In brief, our work introduces a valley-derived field variable – a local magnetic moment distribution in $\vec{r}$ -space, and employs the variable to explore, from the local perspective, valley physics in general and that in inhomogeneous structures in particular. In a nutshell, the work culminates at findings of a variety of distinct local phenomena which are borne out of the inhomogeneity and unveil hidden degrees of freedom in valley physics beyond the global one, including effects of space-dependent external fields and one that breaks the rigid rule of broken inversion symmetry permitting materials with inversion symmetry, gapless single-layer graphene in particular, to be added to the list of family capable of exhibiting nonvanishing magnetic moments. Such findings significantly impact applications. For example, because large flakes of gapless single-layer graphene with good crystallinity can routinely be produced with the exfoliation method [1,2] or the state-of-the-art 2D crystal growth [61–66], critical device fabrications in relevant experiments and applications are considerably eased.

As the concept of local magnetic moments is central to the study, an introductory sketch of the concept is given below including an elucidation of its topological origin and relation to valley magnetic moments. For the sketch, we employ **Figure 1**, which illustrates the valley topology from a symmetry perspective, in the case of gapped single-layer graphene (with band gap = $2\Delta$ ). **Figure 1 (a)** shows the graphene crystal structure, where two types of atomic sites, A and B, are present and alternately occupy hexagonal vertices. Inversion symmetry breaking results from the two types of sites having distinct on-site atomic orbital energy, e.g., $\Delta$ and $-\Delta$, respectively, for the $2p_z$ orbital of carbon. When $\Delta = 0$, the inversion symmetry is restored, and the band gap vanishes as well. The crystal shown in the graph can describe TMDCs too, with the assignment of site A to metal atom, e.g., Mo, W and site B to chalcogen atom pair, e.g., $S_2$, $Se_2$, for example. **Figure 1 (b)** shows the band structure, along with band edge state symmetry with respect to three-fold rotations and mirror reflection about the plane, i.e., symmetry elements of the crystal symmetry group ($C_{3h}$). [67] Depending on how band edge states transform under foregoing symmetry operations, they are classified into E" or A" states, and represented by $(x+iy)z$, $(x-iy)z$, or $z$ to indicate corresponding state symmetry. As shown in the graph, the symmetry varies in the band structure, manifesting a twist between the valleys as well as across the gap. Such twist reduces electron intervalley scattering, protecting valley flavor and enhancing valley lifetime for valley-based applications. More importantly, from the topological standpoint, the symmetry twist is similar to that in a Möbius strip, and signifies a nontrivial topology in the fiber bundle of wave vector ($\vec{k}$) parametrized electron states in Hilbert space with nonvanishing valley Chern numbers and valley magnetic moments.

Important implications for local physics follow from **Figure 1**. As noted in the figure, apart from distinct symmetry, conduction and valence band edge states simultaneously are dictated by site A and B orbitals, respectively, connecting state symmetry to local site probabilities and making the symmetry of a state a distribution in space, with the local state symmetry dependent on the local site probabilities of the state. For a given state, as explained in the work, it ends up at spreading the valley topology derived magnetic moment over the structure in the form of a distribution of 'cell-orbital magnetic moments (COMMs)', with each of which defined on a local hexagon describing local, spin-like rotation on the hexagon. In the case of a homogeneous bulk, the distribution is uniform, with sum total giving the valley magnetic moment. However, in the presence of structural inhomogeneity, COMM generally varies with hexagons and, as such, can go beyond valley magnetic moments to describe local valley physics in $\vec{r}$ -space. In the broad context of topological materials, illustration of such a description also has the following interesting allusion for the study of topologically inhomogeneous systems, namely, that apart from well-known local, $\vec{r}$ -space fields, e.g., Ginzburg-Landau superconducting order parameters [68,69], those directly descending from corresponding topological numbers may be present as well and can provide additional $\vec{r}$-space descriptions to facilitate the study.

Besides the close connection between COMM and topology, our pursuit of a local description in terms of COMM is further supported by the two following arguments. Firstly, as a local magnetic moment, it can interact with external magnetic and electric fields, and manifest local field effects suited to the role of theoretically and experimentally guiding the exploration of local valley physics. Secondly, as shown in the work, COMM provides, independent of the actual state of a structure regarding inversion symmetry, a probability-based measure of local, electron state-dependent inversion symmetry breaking resembling a Ginzburg-Landau symmetry-breaking order parameter, which suggests the feasibility of a Ginzburg-Landau type theoretical formulation of local valley physics in terms of COMM. This work therefore defines the COMM distribution in $\vec{r}$-space as the variable called *valley field* and uses it for the representation of local valley physics. Analytically, a Ginzburg-Landau type framework called '*valley field mechanics*' centering on the valley field and corresponding field equation is developed. Both one and two energy band-based pictures are applied to the development yielding, respectively, valley field equations of variant Schrodinger form suitable for TMDCs and variant Klein-Gordon form suitable for graphene. In terms of such equations, the local aspect of valley physics is explored, including local field effects. A new path – local valley control via space-dependent electric and magnetic fields is opened up. The effect of structural inhomogeneity on valley physics is investigated, both analytically and numerically. A variety of intriguing findings are revealed, as sketched below, with a substantial fraction of them showing characteristics in apparent contradiction to the "global perspective-based expectations and/or constraints".

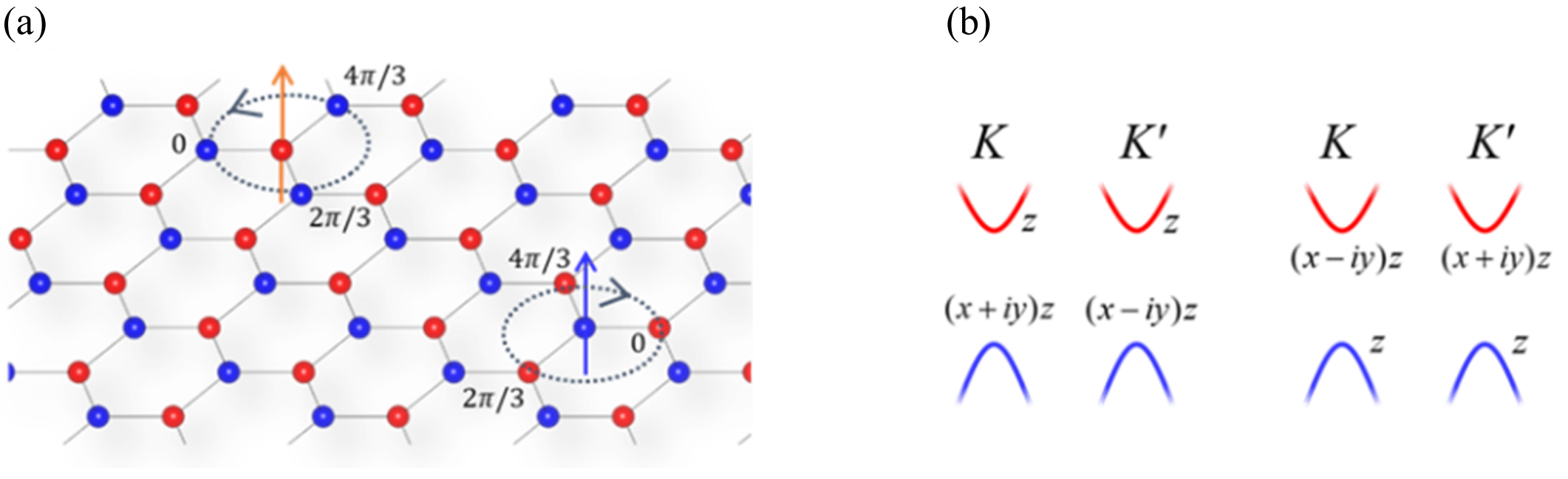

**Figure 1. State symmetry twist and inversion symmetry breaking** Gapped single-layer graphene is used for illustration. **(a)** With on-site 2p$_z$ orbital energy $\varepsilon_A = \Delta$ for A site (red atom) and $\varepsilon_B = -\Delta$ for B site (blue atom), it breaks the inversion (A↔B) symmetry giving a gapped band structure with band gap $= 2\Delta$, conduction and valence band edges respectively located at Δ and -Δ along with conduction (valence) band edge states being A (B) site 2p$_z$ orbital-dominant [3]. Phases of band edge states are shown in the graph and explained in **(b)**. **(b)** Band structure is presented along with band edge state symmetry with respect to the crystal symmetry group C$_{3h}$ symmetry operations, e.g., three-fold rotations and mirror reflection about the plane, in the case where the center of rotation is an A site (left panel) and a B site (right panel). For example, under a rotation about A (B) site, the valence (conduction) band edge state at K $\phi_{K(valence)}$ ($\phi_{K(conduction)}$) transform with $\phi_{K(valence)} \sim (x+iy)z$ ($\phi_{K(conduction)} \sim (x-iy)z$) giving a phase increment $2\pi/3$ ($-2\pi/3$) around the red (blue) out-of-plane arrow as shown in **(a)**. In the terminology of group theory, they are said to belong to the E″ irreducible group representation. On the other hand, under a rotation about B (A) site, the valence (conduction) band edge state at K belong to the A″ irreducible representation and transform with $\phi_{K(valence)} \sim z$ ($\phi_{K(conduction)} \sim z$). Altogether, the state symmetry is twisted across the gap and between the valleys.

**1) Local magnetic moment-external field interactions**

As shown in the work, COMM can couple with space-dependent electrical and magnetic fields giving rise to local valley-orbit and valley Zeeman interactions, respectively.

**2) Breaking of "valley flavor ↔ magnetic moment orientation" correspondence**

In a homogeneous bulk, there a correspondence between the valley flavor and the sign of valley magnetic moment [10,11]. In inhomogeneous structures, however, the study sends an important signal, namely, apart from a space-modulated magnitude, *the sign of local magnetic moment can flip in space*. For magnetic (electric) valley control through the local valley-Zeeman (valley-orbit) interaction, it creates a flexibility, namely, that using local magnetic (electric) fields of opposite signs, with signs correlated with those of local magnetic moments, can achieve the same control.

**3) Disappearing valley magnetic moments**

Valley magnetic moments exist in a homogeneous bulk with broken inversion symmetry, with important physical manifestations. However, the study of valley fields reveals a surprise – suppression or even elimination of sum totals of the fields (i.e., valley magnetic moments) for states near the Dirac point, in zigzag graphene nanoribbons with gap parameter $\Delta \neq 0$. Such a finding apparently violates the global perspective-based expectation.

**4) Nonvanishing magnetic moments and valleytronics in**

**materials with inversion symmetry**

By either symmetry or topological arguments, valley magnetic moments vanish in a homogeneous bulk with inversion symmetry such as gapless graphene ( $\Delta$ = 0). However, as shown in our work, antisymmetric but finite COMM distributions exist in zigzag nanoribbons of gapless graphene. Together with the existence of local valley Zeeman and valley-orbit interactions, it revokes the 'broken inversion symmetry' rule and enables *local valley control-based valleytronics in materials with inversion symmetry*, e.g., gapless single-layer, bilayer graphene etc.

**5) Contrasting material dependence**

The existence (lack) of spin-valley locking in TMDCs (graphene) marks a well-known contrast between the two materials [10,11,23–26]. From the local perspective, valley physics is found to exhibit the following additional material dependence − in TMDCs the valley field always shows a uniform sign in space while in graphene it shows versatile behaviors, including a possible sign flip.

**6) Direct-Indirect band gap control**

The presence of a non-odd potential in Q1D structures is found to have profound effects on band structures. In zigzag graphene ones, such effects are shown to result in the induction of indirect gap near a Dirac point. It implies the feasibility of direct-indirect gap control via an electric potential.

**7) Valley field engineering**

Valley fields in confined structures vary with boundaries, quantization, and types of structures. Therefore, they can be engineered. Such engineering can be integrated with local valley control for versatile applications.

The presentation is organized as follows. In **Sec. II,** we discuss the notion of valley fields, both generically and in the Dirac model of graphene. **Sec. III** presents numerical results of valley fields in various structures, of graphene and TMDCs, with a focus on physical behaviors of the fields. **Sec. IV** turns attention to the framework of valley field mechanics – valley field equations both in the absence and in the presence of external fields, including a description of local valley-Zeeman and local valley-orbit interactions useful for local valley control. **Sec. V** gives conclusion and outlook. The **Appendices** present theories complementary to those given in the main text (in **Appendices A**, **B**, and **D**) and supply mathematical details of derivations (in **Appendix C**). Specifically, **Appendix A** describes the one-band picture-based Schrodinger theory of valley fields. **Appendix B** develops the two-band picture-based Klein-Gordon theory of valley fields, in structures confined by abrupt, asymmetric boundaries. **Appendix C** illustrates the Klein-Gordon theory of valley fields in space-dependent magnetic and electric fields, using structures confined with barriers as examples. **Appendix D** describes numerical results as well as an analytical, perturbation-theoretical treatment of valley fields in the presence of valley mixing.

# II. Valley field: a local concept

Locality and correlation constitute a pair of notions that have marked several decades long, milestone developments in physics. At times they are deemed mutually exclusive, as in the example of epic research in locality principle vs. quantum correlation / nonlocality in multipartite systems, with the EPR paradox [70] and Bell inequalities [71–74] in the early focus of exploration. In the broad interpretation of the notions, however, diverse examples exist where they go hand in hand and provide descriptions complementary to each other, in the condensed matter systems with traditional symmetry-breaking ordered phases [75]. Consider a general, inhomogeneous ferromagnetic state, for instance. At its base is the ensemble of electron angular momenta − spins or atomic orbital ones, of subatomic or atomic length scales. Such angular momenta can interact, align themselves macroscopically breaking the continuous rotational symmetry, and form a magnetic ordered state, where the corresponding order parameter − a local field ($\phi(\vec{r})$) is able to describe the spatially varying ordered state featured by a fluctuation correlation with temperature (T)-dependent characteristic length, e.g., $\xi(T) \propto |T - T_c|^{-1/2}$ in the Ginzburg-Landau $\phi^4$ phenomenology [75]. As will be explained below, the center of this work - COMM distribution and $\phi(\vec{r})$ share quite a few common characteristics.

We define the COMM distribution in $\vec{r}$ -space as the local "valley field" to address both local and correlation aspects of valley physics. **Sec. II-1** starts with a qualitative description of COMM and elucidates its physics by illustrating the connection between COMM and both state symmetry and inversion symmetry breaking, using graphene as an example. **Sec. II-2** quantifies the concept of valley field. In particular, it considers extended states in weakly and smoothly modulated structures of graphene, in the Dirac model, and derive an expression of the field. **Sec. II-3** turns to a generic definition of the field independent of the model and material. **Sec. II-4** provides a discussion of gauge invariance for the definition.

In our discussion below and throughout the work, we follow two conventions when referring to the electron wave vector. In the analytical discussion, which is mostly carried out within the Dirac model, the wave vector is defined relative to the Dirac point (K or K') used in the model. In the discussion of numerical results, which are all obtained with the full-zone tight binding model, the wave vector is defined with respect to the Brillouin zone center ($\Gamma$).

## II-1. The symmetry perspective

In the case of graphene, COMM describes the spin-like, local electron orbital rotation while an electron performs global translation, as shown in **Figure 2**. Consider a near-K state, for example. Generally, it superposes the two components, namely, A site-dominant $\phi_{K(conduction)}$ and B site-dominant $\phi_{K(valence)}$ shown in **Figure 1**. As depicted in **Figure 2**, because the two have distinct E″ and A″ symmetry, they exhibit loop currents of opposite senses and compete. Since each current carries the weight of corresponding local probability, the competition yields COMM $\propto \rho_A - \rho_B$ ( $\rho_{A(B)}$ = electron probability on A(B) site), a net, spin-like local rotation or 'local pseudospin'. Two important pieces of symmetry information are carried in COMM about the competition between E″ and A″ symmetry and between $\phi_{K(conduction)}$ and $\phi_{K(valence)}$, as summarized below.

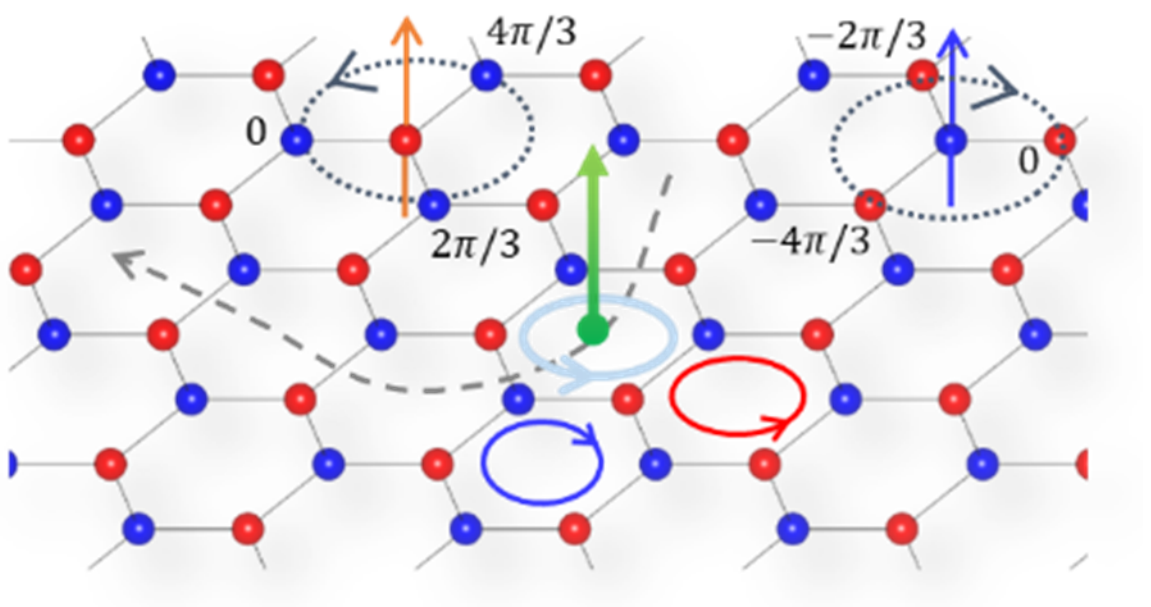

**Figure 2. Cell-orbital magnetic moment** Gapped single-layer graphene is used for illustration, with on-site energy $\varepsilon_A = \Delta$ for A site (red atom) and $\varepsilon_B = -\Delta$ for B site (blue atom). Overall, the electron performs a spin-like, local orbital rotation (light green circle) while simultaneously executing a global translation (grey dashed line). Consider a near-K state, for example. Generally, it superposes the two components – A site-dominant $\phi_{K(conduction)}$ and B site-dominant $\phi_{K(valence)}$, with E″ and A″ symmetry, respectively, as well as corresponding loop currents of opposite senses (orange and blue circles), resulting in the net current $\propto \rho_A - \rho_B$ (light green circle) and corresponding COMM (green, out-of-plane arrow).

### 1) Local dominant state symmetry – E″ or A″

As COMM $\propto \rho_A - \rho_B$, its sign indicates the local dominant state symmetry. Specifically, when COMM changes sign between two regions, it signifies the occurrence in $\vec{r}$ -space of a twist in the dominant symmetry.

### 2) Probability-based inversion symmetry breaking

In analogy to the Ginzburg-Landau order parameter $\phi$, being proportional to '$\rho_A - \rho_B$', COMM serves as the continuous parameter measuring local, '*probability-based*' inversion symmetry breaking, in an electron-state dependent fashion, with $\rho_A - \rho_B = 0$ being the symmetry reference. This measure is independent of the actual state of structure regarding inversion symmetry. Where structural inversion symmetry is present, local probability-based inversion symmetry may actually be broken when $\rho_A - \rho_B \neq 0$.

In summary, on the intra-unit cell scale, a finite COMM signals the existence of a net cell-orbital angular momentum or short-ranged phase correlation among intra-hexagon sites. On the inter-unit cell scale, the COMM distribution is a local field describing spatially varying, *probability-based* inversion symmetry breaking. These foregoing features thus inspire us to introduce the COMM distribution as the "*valley field*" and develop a Ginzburg-Landau type theory to describe both local and correlation aspects of valley physics, with an aim in particular at exploring the physics in inhomogeneous structures.

Throughout the rest of the presentation, we distinguish between the two terms of magnetic moments, namely, COMM and "valley magnetic moments", and reserve the term "valley magnetic moment" for the sum total of COMM distribution, in consistency with the traditional use of the term in the bulk case. Next, we turn to the Dirac model, quantify, and elaborate further on the concept of valley field.

## II-2. Valley field in the Dirac model

A gapped monolayer graphene structure is considered, in the linearized, two-band tight-binding model, also known as Dirac model, where the atomic $2p_z$ orbital per carbon atom is included in the basis set. [3] Generally, the structure is taken to be subject to the modulation of $V(x,y)$, $\Delta(x,y)$ and $B_z(x,y)$ ($V(x,y)$ = electrical potential energy; $2\Delta(x,y)$ = local bulk gap, with $\Delta=\Delta_0$ (constant) $+\delta\Delta$ (modulation); $B_z(x,y)$ = out-of-plane magnetic field with ($A_x(x,y)$, $A_y(x,y)$, 0) the corresponding vector potential, $(x, y)$ = cell position).

Let $F^t = (F_A,\ \ F_B)$ = transposed Dirac two-component wave amplitude on carbon A and B sites, valley index $\tau$ = 1 (-1) for valley K (K'), and $E$ = electron energy. In the Dirac model, $F$ in external fields satisfies the following equation ($\hbar = 1$, $-e$ = 1, and $v_F = 1$ (Fermi velocity) throughout the work) [3,28]:

$$H_{Dirac}F = EF,$$
$$H_{Dirac} = \begin{pmatrix} \Delta + V & -i\partial_x - A_x - \tau\left(\partial_y - iA_y\right) \\ -i\partial_x - A_x + \tau\left(\partial_y + iA_y\right) & -\Delta + V \end{pmatrix}. \quad (1)$$

Consider the relatively simple case of a weakly and smoothly modulated structure in the absence of external fields, where $V = 0$, $B_z = 0$, $|\,\delta\Delta\,| \ll \Delta_0$, and $|\nabla\Delta\cdot(\lambda_x,\lambda_y)| \ll \Delta_0$ ($\lambda_{x(y)}$ = electron characteristic wavelength in the x (y) direction ). From Eqn. (1), one obtains the current distribution $\vec{j}$, with

$$\vec{j} = \vec{j}_f + \vec{j}_m, \quad \vec{j}_m = \nabla\times\vec{m}, \quad (2)$$

where

$$\begin{aligned}\vec{j}_f &= \frac{-iF^\dagger \nabla F + i(\nabla F^\dagger)F}{2E}, \\ \vec{j}_m &= \tau\left(\frac{\partial_y \rho_{diff}}{2E}, -\frac{\partial_x \rho_{diff}}{2E}\right), \\ \vec{m} &= -\frac{\tau \rho_{diff}}{2E}\hat{z}\end{aligned} \qquad (3)$$

( $\rho_{diff}(x,y) \equiv \rho_A(x,y) - \rho_B(x,y)$ , $\rho_{A(B)}(x,y) \equiv |F_{A(B)}(x,y)|^2$ ). From the form of $\vec{j}$ given in Eqn. (2), we are able to identify $\vec{j}_f$ with the free current distribution, $\vec{j}_m$ the magnetization current distribution, and $\vec{m}$ the magnetization distribution, according to magnetostatics [76]. Extension of Eqn. (3) to the case where the modulation is sizable and/or external fields $V$ and $B_z$ are present will be given in **Sec. II-3**. Two important observations based on Eqn. (3) are made below:

i) As $\vec{m} \propto \rho_{diff}$ , it confirms the picture of COMM given in **Figure 2**. Moreover, in the present case, $m$ ( $m \equiv \vec{m}\cdot\hat{z}$ = $-\frac{\tau\rho_{diff}}{2E}$ ) provides a quantitative expression of the valley field, with the sign of $m$ indicating the local, dominant site orbital and state symmetry as mentioned earlier.

ii) For a homogeneous bulk, where $\Delta = \Delta_0$ , the Dirac model gives

$$\begin{aligned}\vec{j}_f &= \frac{\vec{k}}{E}\rho, \\ \vec{j}_m &= 0, \\ m &= \rho\ \mu_{bulk}(E,\Delta_0;\tau), \\ \mu_{bulk}(E,\Delta_0;\tau) &\equiv -\frac{\tau\Delta_0}{2E^2}\end{aligned} \qquad (4)$$

( $\vec{k}$ = wave vector relative to the Dirac point, $\rho(x,y) \equiv \rho_A(x,y) + \rho_B(x,y)$ , $\mu_{bulk}(E,\Delta_0;\tau)$ = valley magnetic moment of the bulk state). $m$ in this case is simply the $\rho$ *-weighted distribution of* $\mu_{bulk}$ in $(x,y)$ space. Eqn. (4) can be alternatively obtained with a topological, valley Berry curvature-based approach [10]. Importantly, it shows the two notable features of $\mu_{bulk}$ established in the field of valleytronics, namely, the correspondence between $\tau$ and sign of $\mu_{bulk}$ , and vanishing $\mu_{bulk}$ in the presence of inversion symmetry ( $\Delta_0$ = 0) [10]. Such features constitute what we call "global perspective-based expectations or constraints".

In the numerical work of this study, magnetic moments in confined structures are expressed in units of "$\mu_B^*$", the magnitude of bulk band edge valley magnetic moment in the well or channel of the structure. In graphene, for example, $\mu_B^* \equiv |\mu_{bulk}(E = \pm\Delta_{channel}, \Delta_{channel};\tau)| = \frac{1}{2|\Delta_{channel}|}$ ( $\Delta_{channel}$ = bulk gap parameter in the channel). When $\Delta_{channel}$ = 0, we use the Bohr magneton $\mu_B$ (5.79× $10^{-5}$ eV/Tesla) in place of $\mu_B^*$.

Next, we leave the Dirac model and develop a generic expression of valley field, which can be applied to TMDCs as well as graphene structures outside the regime of weak modulation.

## II-3. Generic definition

The valley field can interact with a local magnetic field to exhibit a Zeeman energy shift. A model-independent, functional derivative expression of valley field can thus be formulated in terms of the local response of the field to a weak probing magnetic field, as follows:

$$\begin{aligned}m(\vec{r}) &= -\left.\frac{\delta E_{Zeeman_valley}[B_z^{(probe)}(\vec{r})]}{\delta B_z^{(probe)}(\vec{r})}\right|_{B_z^{(probe)}(\vec{r})=0}, \\ E_{Zeeman_valley}[B_z^{(probe)}(\vec{r})] &= -\int m(\vec{r}) B_z^{(probe)}(\vec{r}) d^2 r\end{aligned} \qquad (5)$$

( $E_{Zeeman_valley}$ = valley-Zeeman energy, $B_z^{(probe)}$ = probing magnetic field). Eqn. (5) exploits the physics of local Zeeman interaction "$-m(\vec{r})B_z^{(probe)}(\vec{r})$" to define $m(\vec{r})$ .

The probing field is taken to be a hexagonal (strip) flux in the QD (Q1D) structure as shown in **Figure 3**. In the case of Q1D structure, usage of the strip flux results in $m(y)$ with translational symmetry in the x-direction, consistent with the existence of the symmetry in the structure.

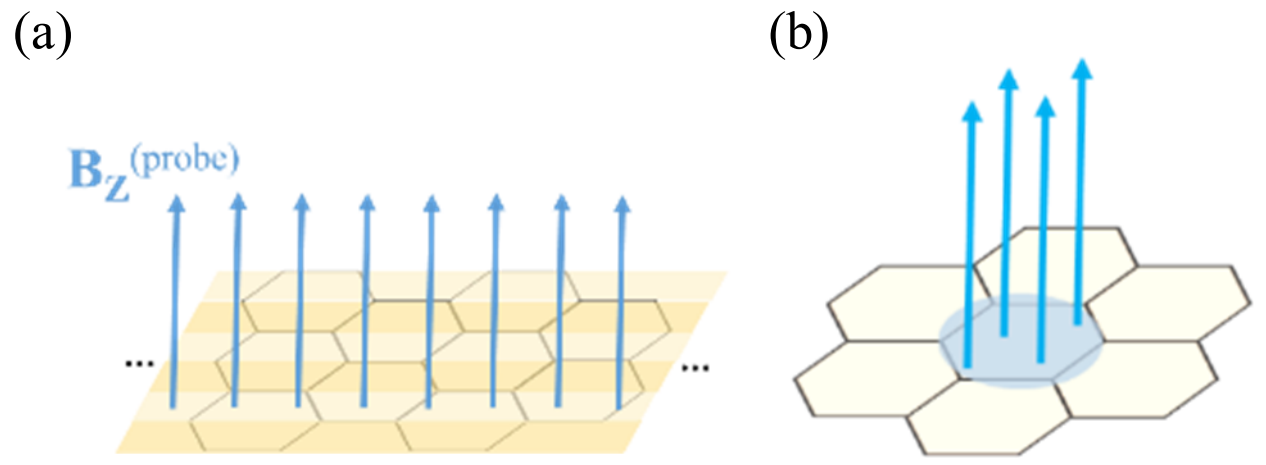

**Figure 3.** $\mathbf{B_z^{(probe)}}$ **(a)** A strip of local, vertical magnetic field in the case of a Q1D structure. **(b)** A hexagonal magnetic flux in the case of a quantum dot.

While $E_{Zeeman_valley}$ in Eqn. (5) sums all the local valley-Zeeman energy '$-m(\vec{r})B_z^{(probe)}(\vec{r})$', it constitutes only a part of the total electron-magnetic field interaction energy. There are other contributions, namely, Landau orbital-magnetic field interaction and non-valley, e.g., spin Zeeman energy as expressed below:

$$E_{Zeeman_valley} = \begin{cases} E_{total} - E_{Zeeman_other} & \text{(QD)} \\ E_{total} - E_{Zeeman_other} - E_{Landau} & \text{(Q1D)} \end{cases}. \qquad (6)$$

Above, $E_{total}$ = total electron-magnetic field interaction energy, $E_{Zeeman_other}$ = non-valley Zeeman energy, and $E_{Landau}$ = $-\left\langle A_x^{(probe)} \right\rangle \int \left\langle j_x \right\rangle dy$ (Landau orbital-magnetic field interaction energy). $A_x^{(probe)}(y)\hat{x}$ = probing vector potential in the asymmetric gauge, and $\langle ... \rangle$ denotes the spatial average of expression inside the bracket. Eqn. (6) provides $E_{Zeeman_valley}$ for Eqn. (5).

Eqns. (5) and (6) together operationally define $m(\vec{r})$ in terms of the local magnetic response, irrespective of the electron energy. As such, it is free of the ambiguity issue in association with the concept of valley flavor for the study of valley physics when electron states move away from Dirac points. Moreover, they facilitate both numerical and analytical studies, as follows. In our numerical study of $m(\vec{r})$, the various energy terms on the right-hand side of Eqn. (6) are obtained with a tight-binding model-based band structure calculation in which the effect of magnetic field $B_z^{(probe)}(y)$ is included via Peierls substitution in the model [77]. Calculations with a finite and with a vanishing $B_z^{(probe)}(y)$ gives $E_{\mathrm{total}}$ and the rest of energy terms. The equations are also critical to the analytical study. For example, in the case of Q1D structures, Eqn. (5) leads to the following analytic expression of magnetization current in the x-direction:

$$\partial_y m(y) = j_x(y) - \left[ \int \left\langle j_x \right\rangle dy \right] \rho(y) \tag{7}$$

independent of external fields. In the case of graphene, the current distribution $j_x(y)$ is given by ($\hbar = 1,\ -e = 1,\ v_F = 1$)

$$j_x(y) = \left( \frac{k_x - A_x(y)}{E - V^{(y)}(y)} \right) \rho(y) + \left( \frac{\tau}{2\left[ E - V^{(y)}(y) \right]} \right) \partial_y \rho_{diff}(y), \tag{8}$$

as can be derived with the Dirac Eqn. (1). Above, $V^{(y)}(y)$ is the electrical potential energy profile in the y-direction. Eqns. (7) and (8) generalize Eqn. (3) to the case where external fields are present, expresses $m(y)$ in terms of probability densities, and provide the key to the analytic treatment of local external field effects presented in **Sec. IV** and **Appendix C**.

We note a few more points below:

i) $m(\vec{r})$ in the 2D bulk is a special case of the Q1D expression above in the wide structure limit.

ii) When applied to graphene of the Dirac model, the generic definition can be shown to recover the valley field expression given in Eqn. (3).

iii) $m(\vec{r})$ defined above is gauge invariant, as explained next.

## II-4. Gauge invariance

$m(\vec{r})$ in Eqn. (5) is gauge invariant, if $E_{Zeeman_valley}$ is gauge invariant. Below, we argue for gauge invariance of the expression given in Eqn. (6) for $E_{Zeeman_valley}$, in Q1D structures first, and QDs next. For simplicity, we exclude *non-valley magnetic moments* such as spin ones from the discussion. In the case of spin magnetic moments, the corresponding spin Zeeman energy is well known to be gauge invariant and the exclusion would therefore not have any effect on the gauge invariance proof of $E_{Zeeman_valley}$.

### Q1D structures

With non-valley magnetic moments excluded, we write $E_{Zeeman_valley}$ = $E_{total}$ − $E_{Landau}$. In principle, $E_{total}$ can be obtained from the following electron-magnetic field interaction energy integral

$$E_{total}[B_z^{(probe)}(y)] \equiv -\int j_x(y) A_x^{(probe)}(y) dy. \tag{9}$$

However, it is obvious that such an integral is generally gauge dependent. A suitable measure is thus required to avoid a subsequent flow of gauge dependence to the difference $E_{total}$ − $E_{Landau}$, which is introduced below.

In both analytical and numerical studies, we use the asymmetric Landau gauge that preserves the lattice translation symmetry in the x-direction. So, the only allowed gauge transformation is the uniform shift: $A_x^{(probe)}(y) \to A_x^{(probe)}(y) + A_0$ ($A_0$ = arbitrary constant). This suggests the following expression for $E_{Landau}$:

$$E_{Landau}[B_z^{(probe)}(y)] \equiv -\left\langle A_x^{(probe)} \right\rangle \int \left\langle j_x \right\rangle dy, \tag{10}$$

with which it yields an $A_0$-independent expression of $E_{Zeeman_valley}$, as can be verified easily.

### Quantum dots

In quantum dots,

$$\begin{aligned} &\vec{j}_f(\text{free current}) = 0, \\ &\vec{j} = \vec{j}_m(\text{magnetization current}) = \nabla \times \vec{m}. \end{aligned} \tag{11}$$

With Eqn. (11), Eqn. (9) becomes

$$E_{total}[B_z^{(probe)}(x, y)] = -\int m(\vec{r}) B_z^{(probe)}(\vec{r}) d^2 r \tag{12}$$

Since the above right-hand side is explicitly a functional of $B_z^{(probe)}(x, y)$, gauge independence of $E_{total}[B_z^{(probe)}(x, y)]$ is ensured. Moreover, it shows $E_{Zeeman-valley}[B_z^{(probe)}(y)] = E_{total}[B_z^{(probe)}(y)]$ in this case, reproducing Eqn. (6) in the case of QDs when non-valley magnetic moments are excluded.

## III. Valley field: physical behaviors

Due to the topological distinction between open and compact spaces, a new realm of valley phenomena beyond the single valley Chern number description or global perspective-based expectations may arise in QD / Q1D structures.

In this section, physical behaviors of valley fields are discussed in confined structures of the following classes, respectively:

**i)** barrier-confined graphene structures,
**ii)** confined TMDC structures,
**iii)** graphene structures confined with abrupt, asymmetric boundaries,

These structures are selected to show 1) unique local phenomena contradicting "global perspective-based expectations", with Classes **i)** and **iii)**; 2) contrast between TMDCs and graphene in local valley physics, with Classes **i)-iii)**, and 3) nontrivial boundary effect, with Class **iii)**.

For insights into the physics, materials and structures considered are relatively ideal and simple. For example, graphene is taken to be single layered and $\Delta$ -modulable. QDs are taken to be square ones, with the armchair (zigzag) axis running in the x (y) direction. Q1D structures are oriented in the x-direction in our convention. Structures discussed in this section are all free from valley mixing. In addition, while edge states are known to exist in structures with abrupt boundaries [3], they are surface properties generally sensitive to chemical treatment and passivation. As such, the discussion of valley fields is focused on "non-edge states" throughout the work.

**Sec. III-1** describes a reference, one-band-based physical picture. **Sec. III-2** presents numerical results.

### III-1. Reference picture

#### One-band picture

As a reference, we introduce the zeroth-order picture based on one-band effective mass approximation for smoothly modulated structures [27]. It writes a near-band-edge state $\psi$ of conduction or valence bands in the form

$$\psi(x,y) \approx f(x,y)\phi_\tau(x,y) \tag{13}$$

($\psi(x,y)$ = total wave function, $f(x,y)$ = slowly varying envelop function, $\phi_\tau(x,y)$ = band-edge Bloch state of valley $\tau$). Eqn. (13) describes the state as being locally given by $\phi_\tau(x,y)$, with an overall amplitude subject to the global modulation of $f(x,y)$. It suggests for the valley field the following corresponding expression:

$$m(x,y) \sim \rho(x,y)\mu_\tau, \tag{14}$$

where $\mu_\tau$ is the valley magnetic moment of bulk band edge state $\phi_\tau(x,y)$ (e.g., $\mu_\tau = \mu_{bulk}(\Delta_0, \Delta_0; \tau)$ in graphene), with an overall magnitude subject to the modulation of probability distribution $\rho(x,y)$ ($\rho(x,y) = |f(x,y)|^2$).

Eqn. (14) describes what we call the one-band picture of $m(x,y)$, which is virtually a local generalization of the homogeneous bulk expression stated in Eqn. (4). It implies relatively simple, monotonous valley fields, with a uniform sign in space, a strong correlation between $m$ and $\rho$, a "$\tau \leftrightarrow$ sign of $m$" correspondence etc., which are well within "global perspective-based expectations" described in Eqn. (4).

Generally, the one-band description works reasonably well when electron states involved are on the band gap energy scale sufficiently near band edges. From such a perspective, the description is generally suitable for TMDCs including confined structures, given their relatively wide band gaps (O(eV)) and heavy effective masses [8,9], as can be verified later in **Sec. III-2**.

#### Beyond the picture

Graphene typically has a relatively narrow gap (O(10-100 meV)) [4–7] and a light effective mass. Therefore, with respect to the band gap scale, electron states involved may easily be relatively away from band edges, mix both conduction and valence band edge states or, equivalently, $2p_z$ orbitals of A and B sites, giving roughly the valley field $m \propto \rho_A - \rho_B$ and invalidating the foregoing one-band description. In the case of confined states, breakdown of the one-band picture may become even more dramatic since, depending on the state energy and structure involved, standing waves on sites A and B may oscillate with distinct phases and wavelengths, giving separate fluctuations in '$\rho_A$' and '$\rho_B$' with a crossover "$\rho_A / \rho_B > 1 \leftrightarrow \rho_A / \rho_B < 1$" in space and a corresponding sign flip in $m$. Overall, a Dirac model-based two-band picture is called for a suitable description of valley physics in graphene.

Two mechanisms listed below strongly differentiate sites A and B and, hence, contribute to the above oscillation contrast:

i) asymmetric boundaries − in the case of a zigzag nanoribbon, the two boundaries terminate at A and B sites, respectively;
ii) on-site atomic orbital energy difference between A and B sites.

Overall, the one-band picture will serve as a guide in **Sec. III-2** to identify bulk-like or global perspective-based behaviors. More importantly, breakdown of the picture will be used to pinpoint modulated structure-specific behaviors that are beyond the reach of single valley Chern number description. Where the breakdown occurs, relatively sophisticated two-band approaches are required in place of the one-band picture. Such approaches include primitive quantum mechanics with the Dirac model based study illustrated in **Sec. II-2** an example, and the two-band-based Klein-Gordon theory of valley fields introduced later in **Sec. IV**.

### III-2. Numerical results

For the numerical work, we apply the tight-binding model of graphene with $2p_z$ orbital per carbon in the atomic orbital basis set [78] and that of TMDCs with $3d\uparrow/\downarrow$ orbitals per metal atom in the set [79,80]. In both models, only nearest neighbor hopping between the orbitals are included.

### i) Barrier-confined graphene structures

**Figure 4** illustrates general features of quantum confined valley fields. A zigzag graphene Q1D structure confined with barriers is considered, with dimensions given by $W_{barrier}$ = 65.2 *a* (barrier width) and $W_{channel}$ = 65.8 *a* (channel width), and bulk gap parameters $\Delta_{barrier}$ = 0.3 eV in the barrier and $\Delta_{channel}$ = 0.1 eV in the channel (*a* = bulk lattice constant). The graph presents valley fields in the case of top and second valence subbands. **(a)** and **(b)** show corresponding valley magnetic moments (sum totals of valley fields) vs. $k_x$ and varying $W_{channel}$, respectively. **(c)** shows the valley field and probability distribution ( $\rho(y)$ ) of the top valence subband state at the Dirac point $k_x \sim -2.10\ a^{-1}$. **(d)** shows $\rho_A(y)$ and $\rho_B(y)$ of the foregoing state. **(e)** shows the valley field and $\rho(y)$ of the second valence subband state at the same Dirac point. **(f)** shows $\rho_A(y)$ and $\rho_B(y)$ of the foregoing state.

Overall, the figure demonstrates a strong correlation between valley field and $\rho(y)$, in the case of the top subband state, as well as breakdown of the correlation when going to deeper valence subbands. In particular, it shows why the sign variation is absent (present) in the case of top (second) subband state: $\rho_A(y)$ and $\rho_B(y)$ have a strong (moderate) contrast in magnitude, and oscillate with different phases and wavelengths, ending up with a uniform (varying) sign in $\rho_A(y)-\rho_B(y)$ and, hence, valley field, too, in the case of top (second) subband state.

**Zigzag graphene Q1D structure**

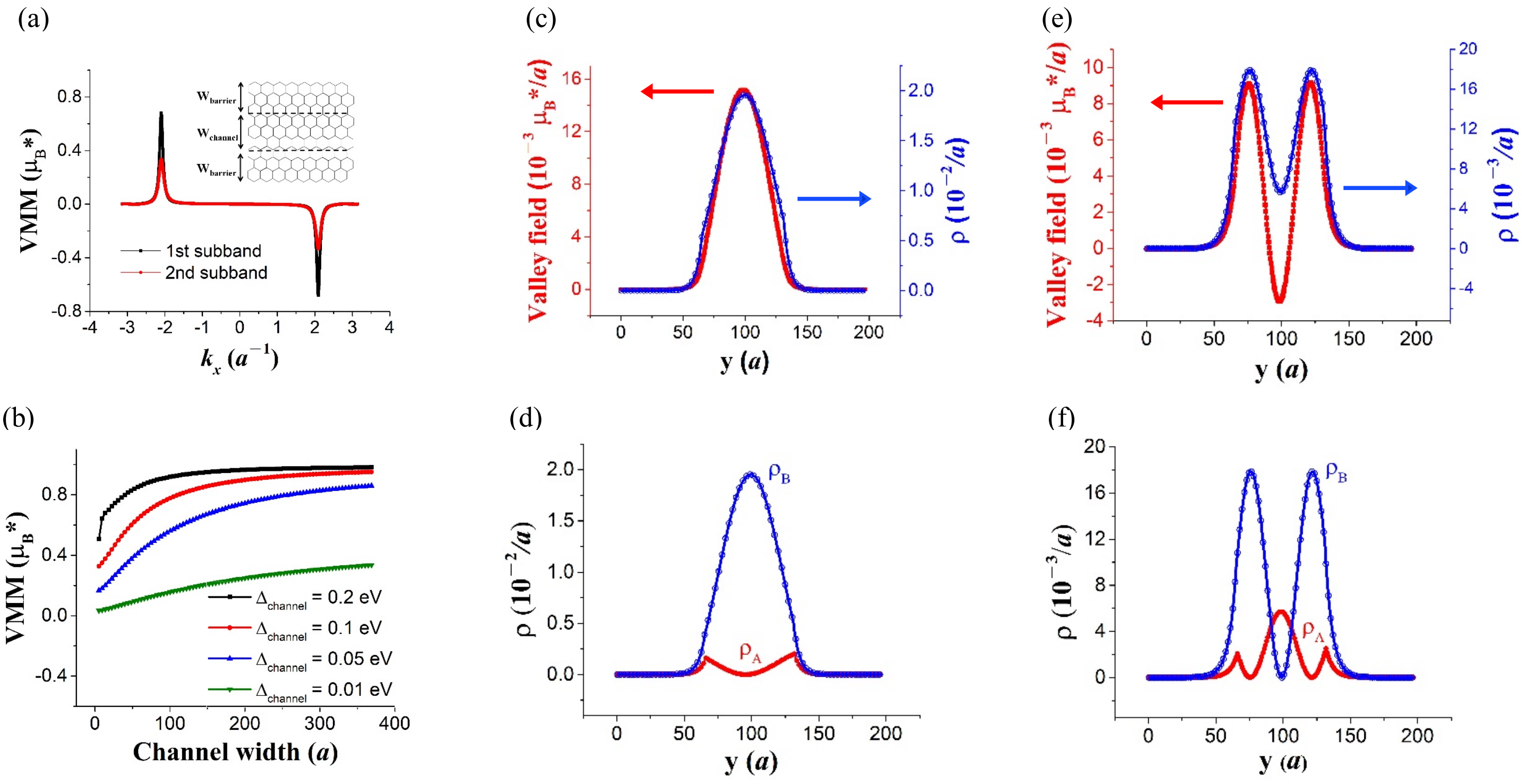

**Figure 4. Valley fields in Zigzag graphene Q1D structure (a)** Valley magnetic moment (VMM) vs. $k_x$, of the first and second valence subbands. VMM decreases in magnitude when moving away from Dirac points or going to deeper valence subbands. **(b)** VMM of the first valence subband state at the Dirac point $k_x \sim -2.10\ a^{-1}$ vs. channel width ($W_{channel}$) as well as gap ( $\Delta_{channel}$ ). It approaches the bulk band edge state limit $\mu_B^*$ when either $W_{channel}$ or $\Delta_{channel}$ increase. **(c)** Valley field and probability distribution ( $\rho(y)$ ) of the top valence subband state at the same Dirac point. **(d)** $\rho_A(y)$ and $\rho_B(y)$ of the state in **(c)**. **(e)** Valley field and $\rho(y)$ of the second valence subband state at the same Dirac point. **(f)** $\rho_A(y)$ and $\rho_B(y)$ of the state in **(e)**. **(c)**-**(f)** shows a strong correlation between the sign and magnitude of valley field and those of $\rho_A(y)-\rho_B(y)$ . **(c)** and **(d)** are well described by the one-band picture. While in **(e)** and **(f)**, with standing waves of A and B sites oscillating in separate phases and wavelengths, the crossover " $\rho_A/\rho_B > 1 \leftrightarrow \rho_A/\rho_B < 1$ " occurs in y along with the manifestation of a sign flip in valley field. A coarse grain averaging is performed in both the valley field and $\rho(y)$ here as well as throughout the work, in the case of Q1D structures, unless noted otherwise.

### ii) TMDC structures

For TMDCs, we focus on valence band states, as they carry an important potential for valley-based applications due to the presence of strong spin-valley-orbital locking and correspondingly induced valley protection [11,81]. For a similar reason, we pick $WSe_2$ as the material. It has, for example, a larger spin-orbit coupling than MoSe2 and a wider K-Γ energy separation than $WS_2$ [82], both of which result in a better protection for the valley flavor in the material.

**Figure 5** shows valley fields in a zigzag $WSe_2$ Q1D structure confined with barriers, with $W_{channel}$ = 21.7 *a* and $W_{barrier}$ = 30.3 *a*. **(a)** shows the subband structure. **(b)** shows VMM vs $k_x$ for each subband. **(c)** shows the valley field and $\rho(y)$ of the top valence subband state at the Dirac point $k_x \sim 2.10\ a^{-1}$. **Figure 6** shows valley fields and probability distributions of the top valence non-edge QD-confined state in two rectangular QDs with different aspect ratios. In **(a)**, $W_x$ = 9.53 *a* and $W_y$ = 13 *a*. The energy of the state is 0.126 eV, with valley magnetic moment (VMM) = 0.76 $\mu_B^*$. In **(b)**, $W_x$ = 19.9 *a* and $W_y$ = 6 *a*. The energy of the state is 0.037 eV, with VMM = 0.62 $\mu_B^*$. $\mu_B^*$ here is the VMM of valence band edge state at Dirac point in bulk $WSe_2$, which is about 4.1 $\mu_B$ [81]. Note that, because of spin-valley-orbital locking, the Kramers two-fold valley degeneracy is protected from the armchair edge induced inter-valley scattering. For demonstration, we choose the valley state with positive VMM to present in the figure.

Overall, in **Figures 5** and **6**, a strong correlation is illustrated between the valley field and probability distribution. Because of the spin-valley-orbital locking protection, the correlation illustrated is expected to be present independent of confining boundary types – abrupt edges or finite barriers, as well as edge orientations - zigzag or armchair ones, as is indeed confirmed in our study (out of which only a selected set of results are presented here).

**Zigzag $WSe_2$ Q1D structure**

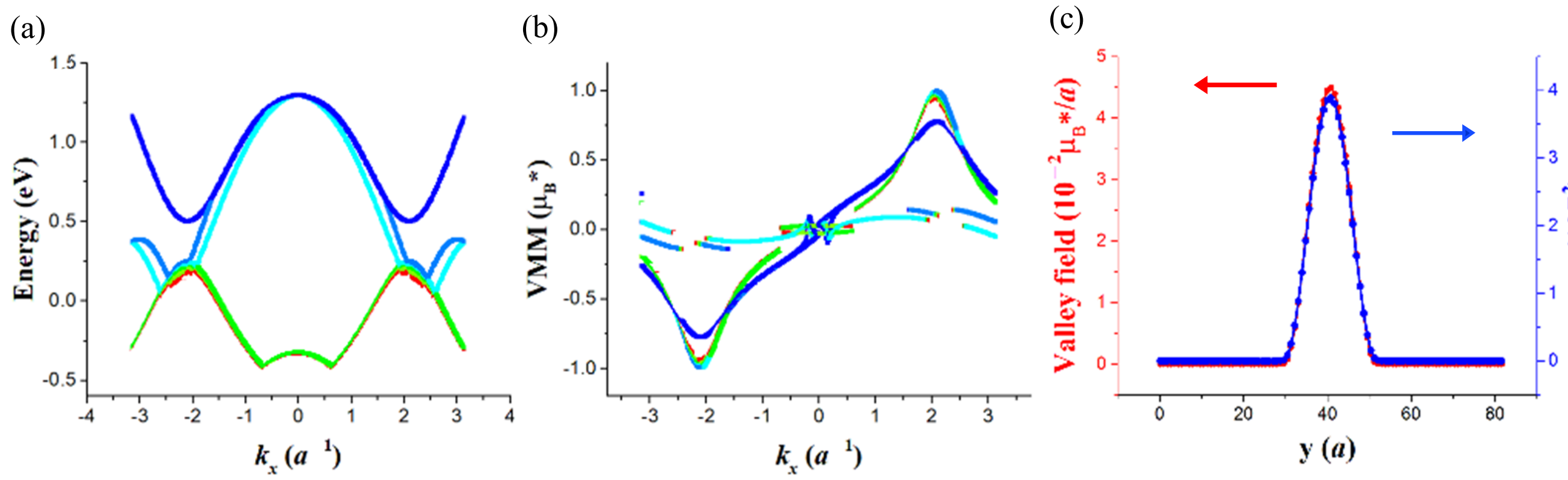

**Figure 5. Valley fields in Zigzag $WSe_2$ Q1D structure** A zigzag $WSe_2$ Q1D structure confined with barriers is considered, with width parameters $W_{channel}$ = 21.7 *a* and $W_{barrier}$ = 30.3 *a*. In order to confine holes to the channel, a negative on-site potential energy (-1 eV) is introduced to form barriers on both sides of the channel. **(a)** Subband structure. **(b)** Valley magnetic moment (VMM) vs $k_x$ for each subband. **(c)** Valley field and $\rho(y)$ of the top valence subband state at the Dirac point $k_x \sim 2.10\ a^{-1}$. The color in **(a)** and **(b)** is used as band index. The barrier potential lowers surface bands into the range of valence subbands and produces band crossings. Due to such crossings, **(b)** shows discontinuity in several colored curves. Overall, the correlation between valley field and $\rho(y)$ is observed to be present independent of confining boundary types – finite barriers here or abrupt edges (not shown), as well as orientations, e.g., zigzag or armchair (not shown) ones.

**$WSe_2$ quantum dots**

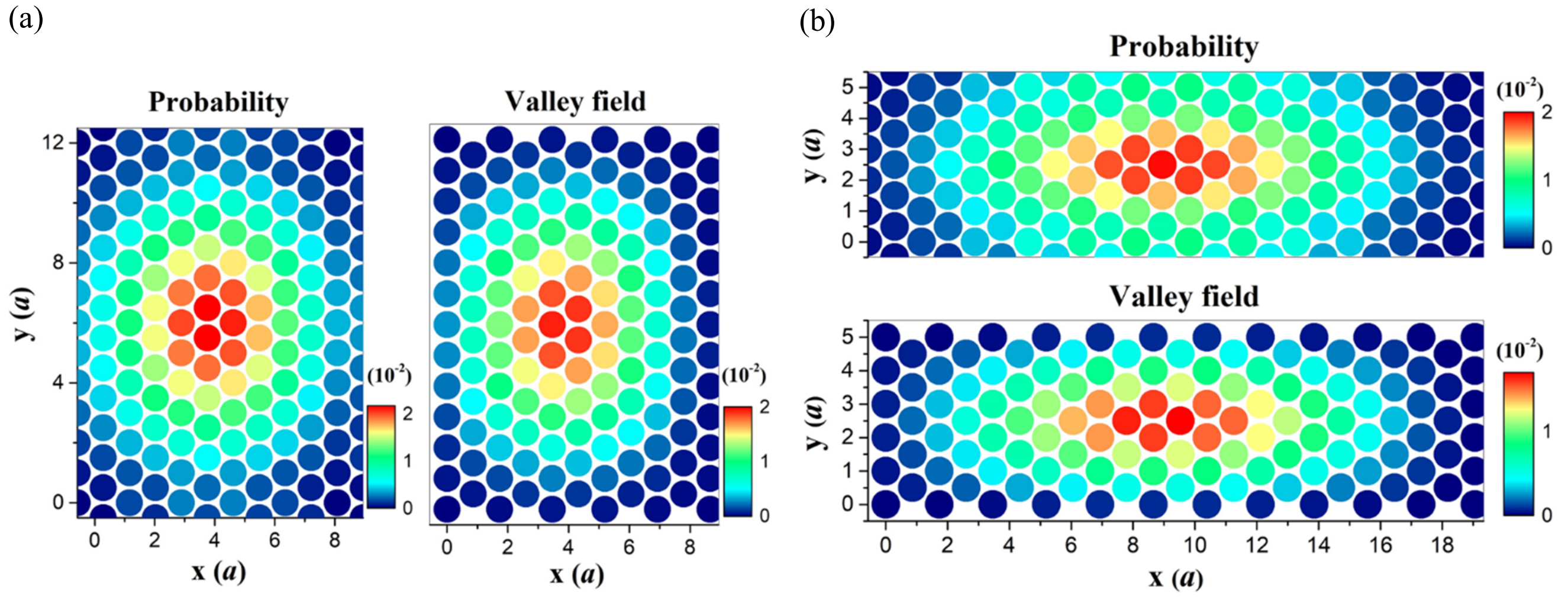

**Figure 6. Valley fields in $WSe_2$ quantum dots** Valley fields (in units of $\mu_B^*$/(hexagon area)) and probability distributions ( $\rho(x,y)$ ) of the top non-edge valence QD-confined state in rectangular $WSe_2$ QDs confined with abrupt boundaries, for two different aspect ratios. In **(a)**, $W_x = 9.53$ $a$ and $W_y = 13$ $a$. The energy of the state is 0.126 eV, with valley magnetic moment (VMM) = 0.76 $\mu_B^*$. In **(b)**, $W_x = 19.9$ $a$ and $W_y = 6$ $a$. The energy of the state is 0.037 eV, with VMM = 0.62 $\mu_B^*$. $\mu_B^*$ is the VMM magnitude of valence band edge state at Dirac point in bulk $WSe_2$, which is about 4.1 $\mu_B$ [81]. Due to the lack of reflection symmetry about the QD vertical center line, probability distributions and valley fields are slightly asymmetric in both **(a)** and **(b)**. Overall, the correlation between valley field and $\rho(x,y)$ is observed to be present independent of confining boundary types – abrupt edges here or finite barriers (not shown).

### iii) Graphene structures confined with abrupt asymmetric boundaries

Zigzag nanoribbon cases are presented to illustrate the profound effect of asymmetric boundaries on topology, in **Figures 7** and **8**, which feature electron states in gapless and gapped zigzag ribbons with gap parameters $\Delta = 0$ eV and $\Delta = 0.1$ eV, respectively, and the same ribbon width W = 65.8 $a$.

In **Figure 7**, **(a)** shows the subband structure. **(b)** shows the valley field accumulated over half width of the ribbon ($VMM_{1/2}$) vs. $k_x$ for each subband. **(c)** depicts valley fields of a few second subband states, at $k_x = -1.88\ a^{-1}$, $-2.10\ a^{-1}$, and $-2.31\ a^{-1}$ in the neighborhood of a Dirac point. **(d)** shows $\rho_A(y)$ and $\rho_B(y)$ of the state at Dirac point ($k_x = -2.10\ a^{-1}$), implying a sign oscillation in $\rho_{diff}(y)$ (i.e., $\rho_A(y) - \rho_B(y)$) and, hence, in corresponding valley field, too, which is consistent with what is shown in **(c)**. Overall, in the gapless case presented in **Figure 7**, valley fields are shown to be always antisymmetric in $y$, thus resulting in vanishing valley magnetic moments independent of $k_x$. However, the field accumulated over half width of the ribbon ($VMM_{1/2}$) is significant and can sometimes exceed 10 $\mu_B$ to provide an access to local valley control, as discussed later in **Sec. IV-2**.

In **Figure 8** of the gapped case, **(a)** shows the top two valence subbands. **(b)** shows valley magnetic moment (VMM) vs. $k_x$ for the two subbands, which reveals vanishing VMMs for edge states (flat part of blue curve in **(a)**), and an overall suppression of VMM near each Dirac point, in the case of second valence subband. **(c)** presents the valley field and $\rho(y)$ of the second subband state at Dirac point. The corresponding $\rho_A(y)$ and $\rho_B(y)$ of the state are shown in **(d)**, which imply a sign oscillation in $\rho_{diff}(y)$ and, hence, in corresponding valley field, too, which is consistent with what is shown in **(c)**.

The two notable observations made above, namely, the existence of finite $VMM_{1/2}$ in the gapless case and suppression of VMM near Dirac points in the gapped case, contradict the naïve picture of valley Chern number-based bulk-like behaviors. In both observations, the sign variation in valley field is right at the center of the phenomena. A clue is given below which connects the sign variation to the nontrivial role played by the asymmetric boundary condition of vanishing site amplitudes, e.g., $F_A(x = W/2) = F_B(x = -W/2) = 0$ [38]. As this condition effectively boosts up the on-site energy of A (B) site on the boundary y = $W$/2 (- $W$/2) to infinity, it introduces in the y-direction *a twist in on-site energy* and consequently in '$\rho_A - \rho_B$', too, resulting in antisymmetric or nearly antisymmetric valley fields. From the theoretical perspective, such twist has a nontrivial effect on valley physics. Apart from the clean or nearly clean elimination of valley magnetic moments shown here, **Appendix B** shows that an intriguing pseudo vector potential parameter "$A_x^{(BC)}(\tau)$" is induced by the boundary asymmetry and shifts the subband edge. The Appendix also supplies an analytical proof of exact antisymmetry in the valley field of second subband edge state.

**Gapless zigzag graphene nanoribbon**

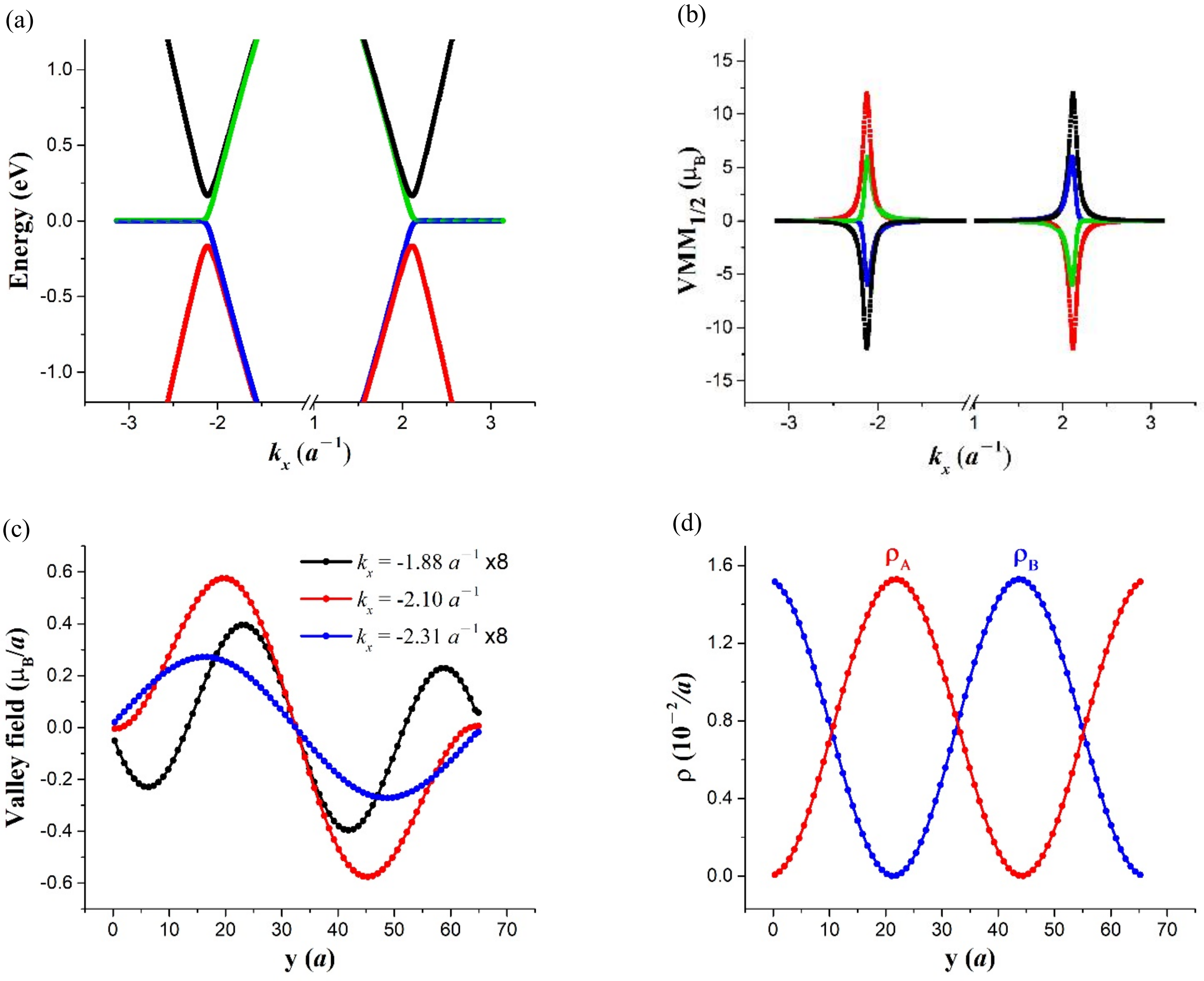

**Figure 7. Vanishing valley magnetic moments (VMMs) but finite valley fields in gapless zigzag graphene nanoribbon (ZGNR)** A gapless ($\Delta$ = 0 eV) ZGNR with W= 65.8 *a* is considered. **(a)** Subbands. **(b)** Valley field accumulated over half width of the ribbon ($VMM_{1/2}$) vs. $k_x$ for each subband. The color in **(a)** and **(b)** is used as band index. While VMM always vanishes independent of the state, $VMM_{1/2}$ can sometimes exceed 10 $\mu_B$. **(c)** depicts nontrivial, antisymmetric valley fields of a few second subband states in the neighborhood of Dirac point ($k_x$ = -2.10 $a^{-1}$). **(d)** shows $\rho_A(y)$ and $\rho_B(y)$ of the state at Dirac point in **(c)**. Overall, a sign oscillation exists in corresponding $\rho_{diff}(y)$ and valley field.

**Gapped zigzag graphene nanoribbon**

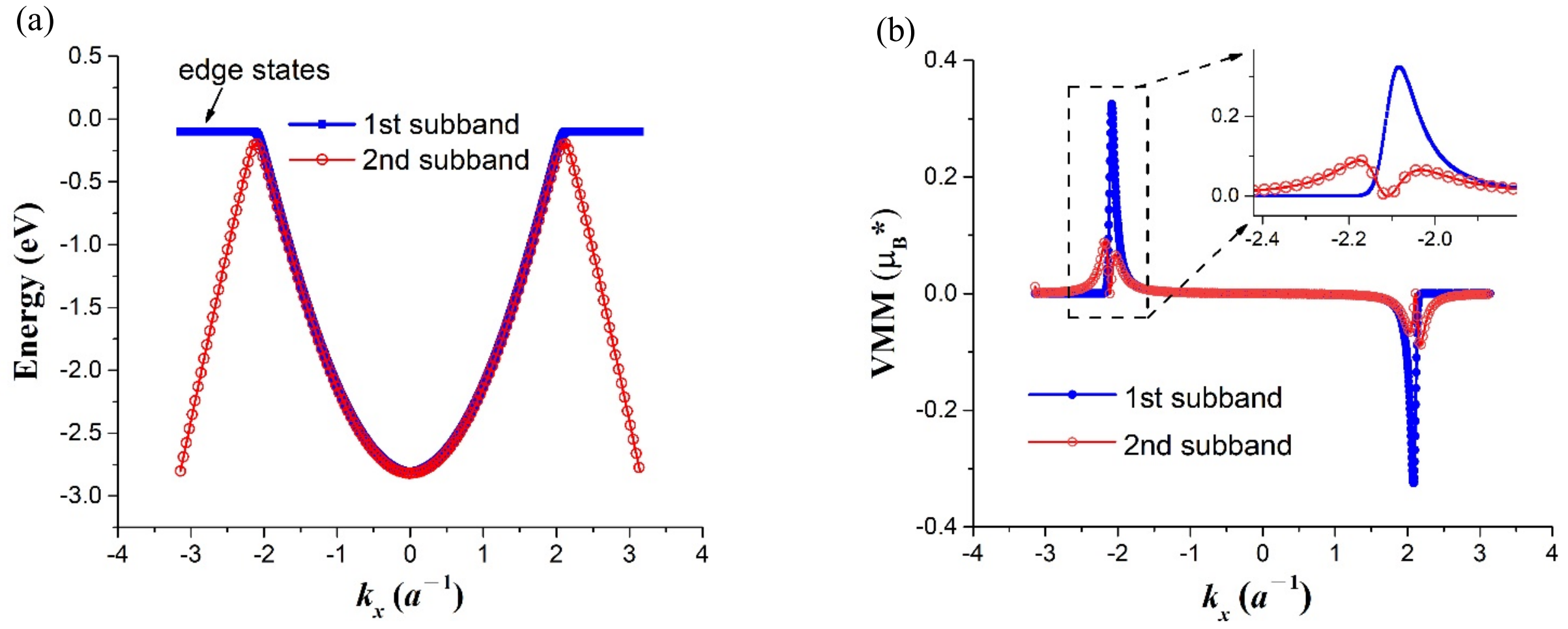

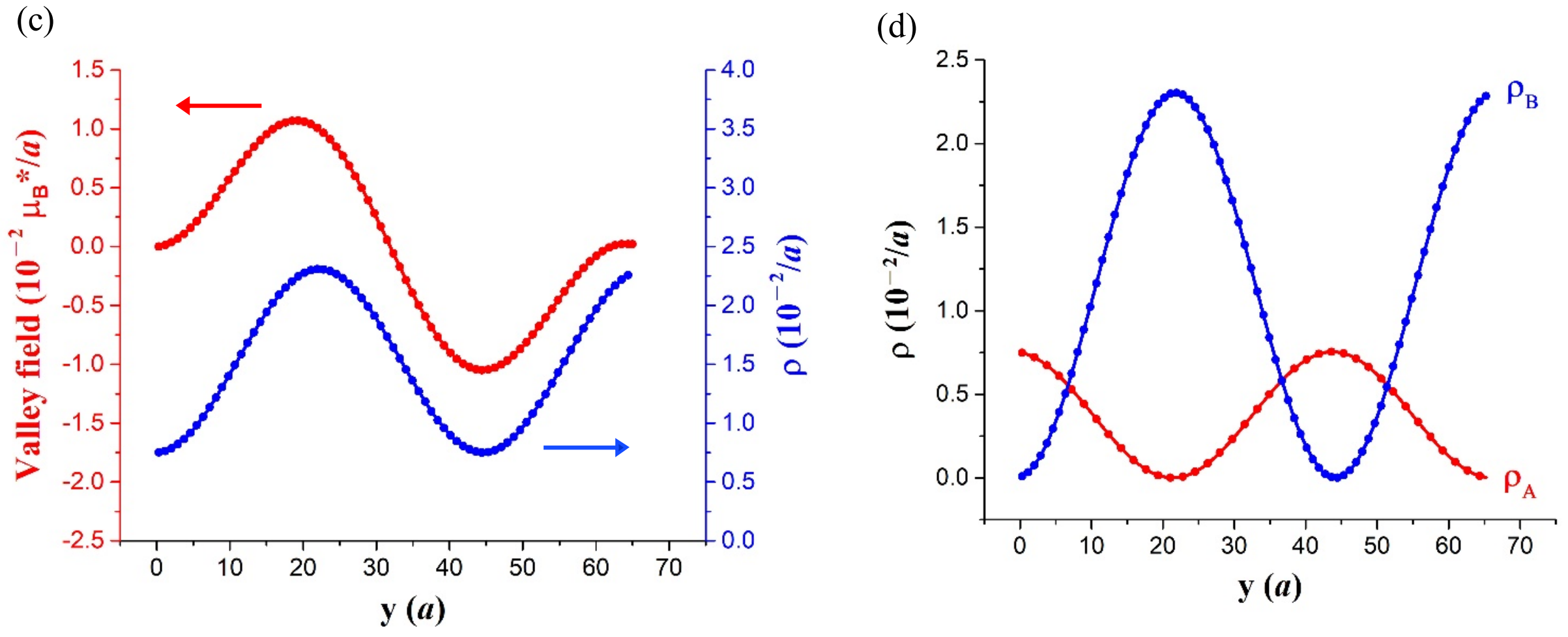

**Figure 8. Valley magnetic moments (VMMs) and valley fields in gapped zigzag graphene nanoribbon (ZGNR)** Gapped ZGNR with abrupt boundaries is considered, with width W = 65.8 *a* and gap parameter Δ = 0.1 eV. **(a)** Top two valence subbands. **(b)** VMM vs $k_x$ for bands in **(a)**, which reveals vanishing VMMs for edge states (flat part of blue curve in **(a)**), and an overall suppression of VMM near each Dirac point ($k_x$ = +/- 2.10 $a^{-1}$), in the case of second valence subband. **(c)** presents valley field and ρ(y) of the second subband state at Dirac point ($k_x$ = -2.10 $a^{-1}$), with corresponding $\rho_A$(y) and $\rho_B$(y) shown in **(d)**.

## IV. Valley field mechanics: field equations and local magnetic / electric field effects

Local valley phenomena can often be analytically studied for insights. For such studies, this section presents "valley field mechanics" centering on valley field equations. Based on the equations, effects of space-dependent external electric and magnetic fields are discussed, with important implications for local valley control via external fields.

The valley field (denoted as $m$ below) is governed by a Ginzburg-Landau type field equation. Depending on the approximations involved, the equation can be formulated in various analytic forms. Two relatively simple forms of the equation, namely, one-band picture based *variant Schrodinger form* and two-band picture based *Klein-Gordon form* are illustrated in this work. As mentioned in **Sec. III**, the one-band picture is suitable for the relatively monotonous valley physics in TMDCs while the two-band picture is needed for the relatively versatile valley physics in graphene.

As valley fields vary with materials and structures, we select the following case for presentation, namely, graphene Q1D structures confined with barriers. **Sec. IV-1** is focused on discussing the *Klein-Gordon form* in the case, leaving a description of the Schrodinger form to **Appendix A**. External field effects in the structures are discussed in **Sec. IV-2**. **Sec. IV-3** remarks on the general nature of valley field mechanics to close the section. Mathematical details for **Secs. IV-1** and **IV-2** are provided in **Appendix C**.

### IV-1. The variant Klein-Gordon theory

The Klein-Gordon theory summarized below provides a Dirac two-band based description of valley fields in the absence of external fields.

#### Graphene Q1D structures confined with barriers

The theory presented applies to barrier-confined graphene Q1D structures, both zigzag and armchair oriented. We take each barrier to be semi-infinite to keep electrons away from the boundary and ignore the effect of abrupt asymmetric boundaries in the zigzag case. In the armchair case, due to the barrier blockade, the intervalley coupling comes only from the channel-barrier interface scattering, which is limited in strength and also ignored below. In this case, the theory presented is to be regarded as the zeroth-order theory. Similar to **Sec. III**, this section primarily addresses valley fields in the limit of vanishing intervalley coupling, leaving the analytical treatment of limited valley mixing effect to **Appendix D**.

The structures considered are taken to be subject to a gap modulation $\Delta^{(y)}(y)$, with $\Delta^{(y)}(y)$ effecting a semi-infinite barrier confinement. We also exclude external fields. As shown in **Appendix C**, the Dirac model leads to the following valley field equation:

$$H_{KL} m(y) = E^2 m(y),$$

$$H_{KL} \equiv -\frac{1}{4}\partial_y{}^2 + k_x{}^2 + \Delta^{(y)}(y) \int_{-\infty}^{y} \Delta^{(y)}(y)\partial_y dy \qquad (15)$$

along with the boundary condition

$$m(\pm\infty) = 0 \qquad (16)$$

Above, $H_{KL}$ is a Klein-Gordon type operator.

The theory expressed by Eqns. (15) and (16) excludes the presence of any electrical modulation other than $\Delta^{(y)}(y)$ or magnetic vector potential. Being external field-free, the theory serves as the zeroth-order description when studying effects of external fields.

## IV-2. Effects of external fields

Effects of space-dependent electric and magnetic fields are discussed below, with focus on two aspects:

i) *local valley-external field interactions*, for valley control via external fields;
ii) *linear valley field response* to external fields.

An illustration is given again in the case of graphene Q1D structures, which are taken to be subject to the electrical modulation $V^{(y)}(y)$ and / or magnetic vector potential $A_x(y)\hat{x}$.

### Local valley-external field interaction

For a homogeneous bulk in the presence of a uniform, out-of-plane magnetic field, e.g., $B_z$, the electron valley magnetic moment $\mu_{bulk}$ interacts with $B_z$ showing the valley Zeeman energy "$-B_z\mu_{bulk}$" [10,28]. This fact has been exploited in **Sec. II-3** when introducing the generic definition of local magnetic moment *m*, using the generalized local expression "$-B_z^{(probe)}(y)m(y)$" for the interaction between *m* and the probing magnetic field $B_z^{(probe)}(y)$ both of which vary in space.

Similarly, in the presence of a uniform, in-plane, transverse electric field $\mathcal{E}_y$, $\mu_{bulk}$ interacts with $\mathcal{E}_y$ showing the valley-orbit interaction energy "$\frac{k_x}{\Delta_0}\mathcal{E}_y\mu_{bulk}$" – a Rashba term in the electron energy, in the bulk graphene case [28]. This fact may also be exploited to define the local magnetic moment, for example, using the straightforward extension "$\frac{k_x}{\Delta_0}\mathcal{E}_y^{(probe)}(y)m(y)$" for the interaction between *m* and the probing electric field $\mathcal{E}_y^{(probe)}(y)$ both of which vary in space.

More importantly, the above discussion leads to the expectation of existence of *local valley – external field interactions* in the forms given by the two foregoing generalized expressions. Such expectation roughly agrees with the rigorous result, as follows. Let $m^{(0)}(y)$ and $E^{(0)}$ be the field-free valley field and electron state energy $E^{(0)}$, e.g., the solution to Eqns. (15) and (16). As shown in **Appendix C**, in the linear response regime, the corresponding local valley-external field interaction energy is given by

$$E_{valley-field} \approx \int_{-\infty}^{\infty} \frac{k_x}{E^{(0)}} \partial_y V^{(y)} m^{(0)}(y)dy - \int_{-\infty}^{\infty} B_z(y)m^{(0)}(y)dy \quad . \tag{17}$$

Eqn. (17) provides explicit, rigorous expressions of local valley-external field interactions. In particular, it gives

$$\frac{k_x}{E^{(0)}}\partial_y V^{(y)} m^{(0)}(y) \tag{18}$$

for the *local valley-orbit interaction* due to electric force (e.g., $-\partial_y V^{(y)}$), and

$$-B_z(y)m^{(0)}(y) \tag{19}$$

for the *local valley-Zeeman interaction* due to magnetic field $B_z(y)$. Such interactions serve as useful *mechanisms* for *local valley control* via space-dependent electric / magnetic fields. Note that in the low energy limit where $E^{(0)} \to \Delta_0$, Eqn. (18) is exactly the local generalization of the expression "$\frac{k_x}{\Delta_0}\mathcal{E}_y\mu_{bulk}$" given earlier in the bulk case to an inhomogeneous structure.

Numerical examples are presented in **Figure 9** for local electric effects and **Figure 10** for local magnetic effects.

In **Figure 9**, a zigzag graphene nanoribbon with vanishing bulk gap is considered. It shows two contrasting subband structures − one with a direct gap when the structure is electric field free and the other with an indirect gap when the structure is subject to a symmetric potential $V^{(control)}(y)$. In order to appreciate the unique local electric effect induced by $V^{(control)}(y)$ here, we turn to a homogeneous bulk below.

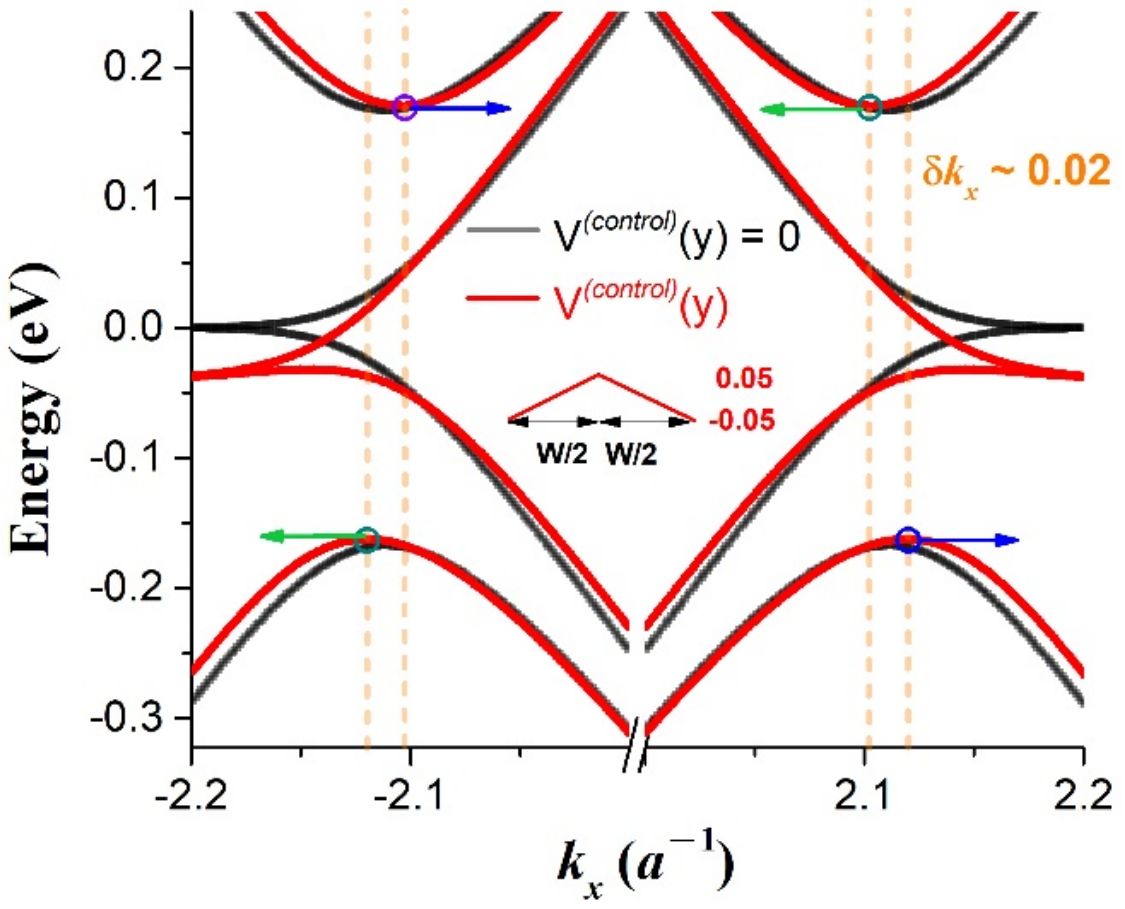

**Figure 9. Local electric effects** The same gapless graphene nanoribbon specified in **Figure 7** is considered. We refer to the gap between the second valence and second conduction subbands. Plotted in the graph are two contrasting subband structures − the black one with a direct gap when the structure is electric field free and the red one with an indirect gap when the structure is subject to a symmetric potential $V^{(control)}(y)$ which varies linearly between $\pm 0.05\ eV$. Due to the local valley-orbit interaction $V^{(control)}(y)$ induces band edge shifts in $k_x$, with opposite signs for the two valleys as well as for conduction and valence subbands, resulting in an indirect gap with $\delta k_x \sim 0.02\ a^{-1}$ (conduction-valence band edge wave vector difference).

Generally, in the bulk case, a simple linear, antisymmetric potential $V^{(control)}(y)$ would suffice to produce notable effects on energy bands. As the corresponding Rashba term "$\frac{k_x}{\Delta_0}\mathcal{E}_y\mu_{bulk}$" ($\mathcal{E}_y = -\partial_y V^{(control)}$) in the case is linear in both $k_x$ and $\tau$, it would displace energy bands in a valley contrasting way, resulting in the so called "valley Rashba splitting" of bands, useful for device applications [34]. However, in the case of gapless graphene ($\Delta = 0$), since $\mu_{bulk}$ vanishes, the splitting disappears, too.

On the other hand, in the ZGNR considered in **Figure 9**, despite $\Delta = 0$, the Rashba splitting is reinstated due to two local effects. Firstly, the field-free valley fields involved are antisymmetric but nonvanishing, as shown earlier in **Figure 7 (c)**. Secondly, with $V^{(\text{control})}(y)$ being symmetric, it results in a finite valley-orbit interaction energy integral $\int_{-\infty}^{\infty} \frac{k_x}{E^{(0)}} \partial_y V^{(\text{control})} m^{(0)}(y) dy$. Indeed, as shown in **Figure 9**, finite, Rashba-type band edge shifts in $k_x$ occur with opposite signs for the two valleys. Moreover, they occur with opposite signs too between conduction and valence subbands, resulting in an indirect gap with finite conduction-valence band edge wave vector difference $\delta k_x \sim 0.02\ a^{-1}$. A similar gap alteration can be shown to occur in zigzag nanoribbons of gapped graphene, too.

We make a note in regard to the direct-indirect gap transformation illustrated. In view that the many-electron Hartree interaction may supply the required non-odd potential, it leads to the conjecture of zigzag graphene nanoribbons being intrinsically indirect-gapped.

Now we turn to **Figure 10**. **Figure 10 (a)** shows the subbands when a *local* magnetic field ($B_z$) is applied to the lower half ribbon. While the involved valley fields are antisymmetric with vanishing sum totals, the total valley-Zeeman interaction energy due to $B_z$ is finite, breaking valley degeneracy and leading to valley splitting. **Figure 10 (b)** shows the subbands when a *uniform* magnetic field ($B_z$) is applied. The total valley-Zeeman interaction vanishes so $B_z$ only introduces an energy shift due to the Landau orbital quantization, thus preserving the valley degeneracy.

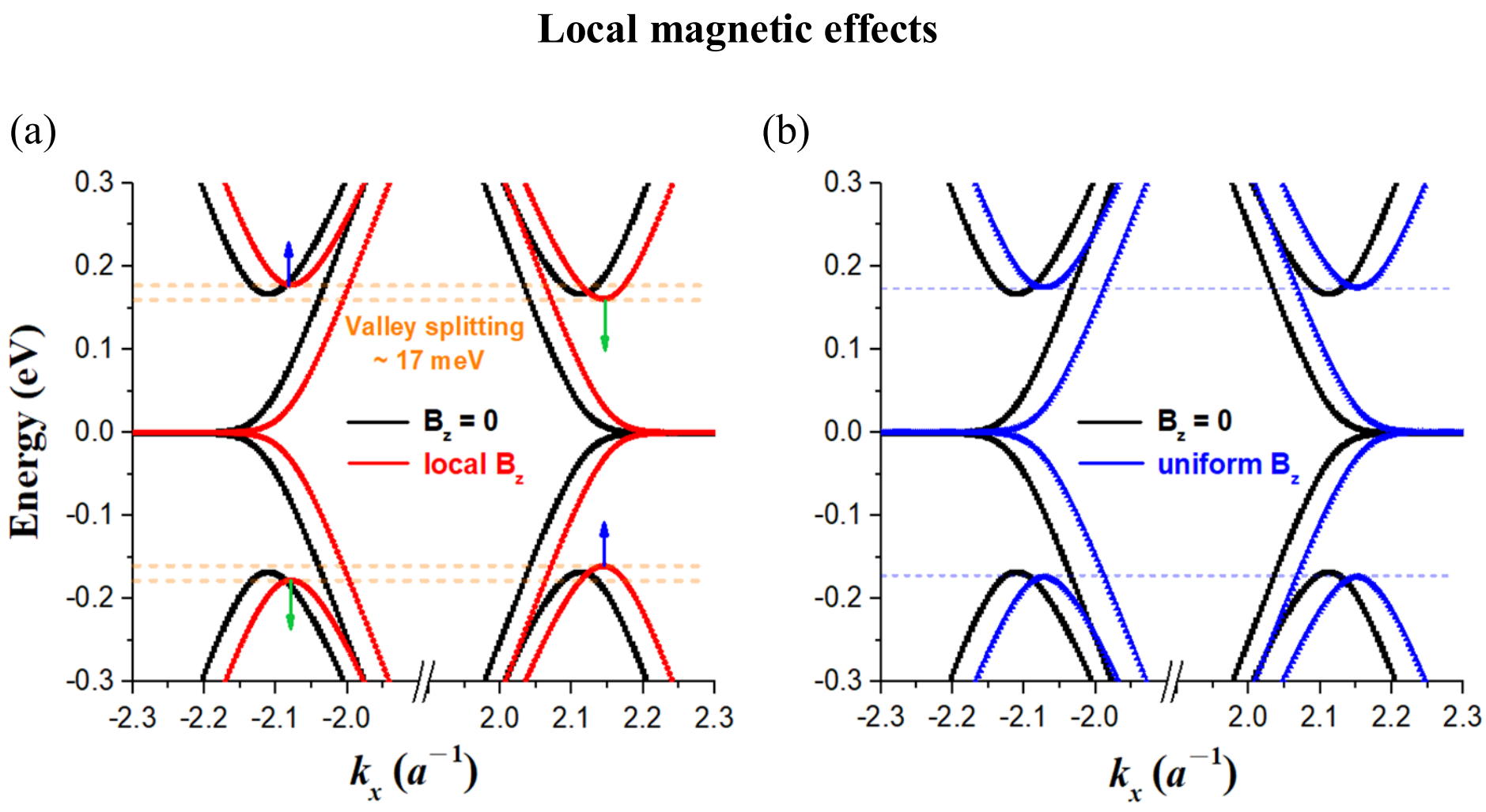

**Figure 10. Local magnetic effects** The same gapless graphene nanoribbon specified in **Figure 7** is considered. **(a)** The subbands when a *local* magnetic field ($B_z$) is applied to the lower half ribbon. It breaks valley degeneracy and leads to valley splitting. **(b)** The subbands when a *uniform* magnetic field ($B_z$) is applied. It only introduces an energy shift due to the Landau orbital quantization, thus preserving the valley degeneracy. $\mu_B B_z = 1$ meV is used in both **(a)** and **(b)**.

Overall, **Figures 9 and 10** send an important message – local valley physics expands flexibility and feasibility in both materials and valley control in valleytronics.

**Local linear response: $m^{(1)}(y)$**

We write $m(y) \approx m^{(0)}(y) + m^{(1)}(y)$ with $m^{(1)}(y)$ the linear response to external fields. As shown in **Appendix C**, a Klein-Gordon valley-field equation with a source term can be formulated for $m^{(1)}(y)$ in the linear response regime:

$$\left[H_{KL} - E^2\right] m^{(1)}(y) = s^{(1)}(y; V^{(y)}, A_x), \tag{20}$$

where the source $s^{(1)}(y; V^{(y)}, A_x)$ is linear in the external fields as given in the Appendix.

Eqn. (20) is derived in the Appendix for the transverse field configuration, where an electric potential energy $V^{(y)}(y)$ and vector potential $A_x(y)$ are present. Such configuration is expected to have implications relevant to three-terminal device based valleytronic signal processing, where the 'valley transconductance' given by the ratio "$m^{(1)}(y)$ / transverse field" would play an important device figure of merit.

## IV-3. Nature of the mechanics

As a theoretical framework, valley field mechanics is featured by the following characteristics.

**1) Intermediate-level quantum description**

It provides a description of valley physics which interpolates between the global valley flavor and the primitive, site-resolved wave mechanics.

**2) Space-dependent description for valley topology**

The observable 'valley field' transforms the $\vec{k}$ -space, valley Chern number-based description to a $\vec{r}$ -space one. It depicts the state symmetry distribution, including possible twists, in $\vec{r}$ -space to suit a general, space-dependent situation. In the homogeneous bulk case, the valley field can be integrated to yield the valley magnetic moment and reflect the corresponding valley Chern number as well.

**3) Normal mode mechanics**

In the graphene case, an interpretation of 'normal mode mechanics' applies to the mechanics. In a sense, { $\rho_A - \rho_B$ , $\rho_A + \rho_B$ } form a set of 'normal mode variables' for the electronic motion, and offer, in comparison to the naive, site-based variables { $\rho_A , \rho_B$ }, a relatively intuitive picture, with ' $\rho_A - \rho_B$ ' describing the *intra-cell orbital motion* and ' $\rho_A + \rho_B$ ' – the probability distribution as a function of cell position describing the global, *inter-cell translational motion*. The mechanics is focused on the 'intra-cell normal mode'.

## V. Conclusion and outlook

In conclusion, valleytronics in 2D materials is rooted in the existence of global valley flavor but extends far out to the rich dimension of local physics, where inhomogeneity in space can come into play and enrich the physics with distinct local valley phenomena not dictated by the traditional perspective of valley topology based on valley Chern numbers of homogeneous bulks or associated global observables of valley magnetic moments.

In order to explore the local dimension, a Ginzburg-Landau symmetry-breaking order parameter type field – valley field has been introduced and served as an important vehicle to address in the presence of inhomogeneity degrees of freedom beyond the valley magnetic moment. It has the interpretation of local cell-orbital magnetic moment and is operationally defined in terms of local magnetic response irrespective of electron state energy, which is free from the ambiguity issue encountered in a valley flavor-based approach to valley physics, of defining non-band-edge state's valley flavor. In graphene, the moment roughly scales with the local site probability difference $\rho_A$ - $\rho_B$, giving the breaking of local probability-based inversion symmetry as the condition for nonvanishing local magnetic moments. The field variable introduced is also application-suited as it is directly linked to local valley-external field effects. For analytical studies, the Ginzburg-Landau type framework of valley field mechanics has been developed comprising valley field equations of variant Schrodinger or Klein-Gordon forms, and applied to the local linear response of valley fields yielding local valley-Zeeman and local valley-orbit-interaction effects, which are critical to the local valley control via space-dependent magnetic and electric fields.

The study has revealed a spectrum of intriguing local valley phenomena, with quite a few profound twists with respect to global perspective-based expectations and/or constraints which evidence the degrees of freedom beyond the global one, such as

– breaking of 'valley flavor ↔ magnetic moment orientation' correspondence,

– lifting of 'inversion symmetry breaking' condition for existence of magnetic moments,

– suppression or even elimination of valley magnetic moments, for near-Dirac point states in gapped graphene structures.

By revoking such constraints, the foregoing findings have expanded the flexibility in valleytronics. For example, gapless, single-layer graphene, a material with inversion symmetry can now be added to the list for magnetic moment-based experiments or applications. Another example is given by the valley field sign flip in space in confined structures, which has the implication for flexible valley control, namely, local magnetic (electrical) fields of opposite signs with signs locally correlated to those of the sign-varying valley field may produce the same valley-Zeeman splitting (valley-orbit interaction) and effect the same magnetic (electric) valley control. Apart from the foregoing findings, the study has also yielded additional insights into the material dependence of valley physics, e.g., relatively versatile valley fields in graphene vs. relatively monotonous ones in TMDCs, besides the well-known spin-valley locking contrast.

Last, the diverse local valley phenomena shown here suggest the attractive direction of valley field engineering, e.g., design and search for quantum structures to tailor valley fields via confinement, defects, boundaries, dopants, constituent materials — single- / multi- layer graphene (with [20–22,83] / without twists), single- / multi (homo or hetero)- layer transition metal dichalcogenides with parallel / antiparallel stackings (with / without twists) [25,26,47,48,84–87], etc. to suit applications.

## Acknowledgement

We thank Prof. Mei-Yin Chou for various discussions, and Yen-Ju Lin for technical support in numerical calculations. We acknowledge the financial support of MoST, ROC through Contract No. MOST 109-2811-M-007-561. F.-W. C. acknowledges partial support from Academia Sinica.

**Table of important notations**

| Notation | Definition |
|---|---|
| $(A_x, A_y)$ | Vector potential in the 2D layer |
| $A_x^{(BC)}$ | Pseudo vector potential due to asymmetric boundary conditions in a Q1D structure |
| $A_x^{(probe)}$ | Probing vector potential in a Q1D structure |
| $a$ | Lattice constant |
| $B_z$ | Magnetic field out of the 2D layer |
| $B_z^{(probe)}$ | Probing magnetic field out of the 2D layer |
| $E$ | Electron state energy |
| $E^{(0)}$ | Electron state energy in the absence of external fields |
| $E_{total}$ | Total electron-magnetic field interaction energy |
| $E_{Zeeman_valley}$ | Valley-Zeeman energy |
| $E_{Zeeman_other}$ | Non-valley Zeeman energy |
| $E_{Landau}$ | Landau orbital-magnetic field interaction energy |
| $E_{valley\text{-}field}$ | Valley field-external field interaction energy |
| $\mathcal{E}_y$ | Electric field in the y-direction |
| $\mathcal{E}_y^{(probe)}$ | Probing electric field in the y-direction |
| $(F_A, F_B)$ | Transposed wave amplitude on A and B sites in the Dirac model of graphene |
| $f$ | Envelop wave function in the effective-mass description |
| $H_{Dirac}$ | Hamiltonian in the Dirac model of graphene |
| $H_{KL}$ | Klein-Gordon type operator in the valley field equation |
| $H_S$ | Schrodinger type operator in the valley field equation |
| $\vec{j}$ , $(j_x, j_y)$ | Current density |
| $\vec{j}_f$ | Free current density |
| $\vec{j}_m$ | Magnetization current density |
| $j_x^{(0)}$ | $j_x$ in the absence of external fields |
| $\vec{k}$ , $(k_x, k_y)$ | Electron wave vector, defined with respect to a Dirac point in the analytical study and the Brillouin zone center (Γ point) in the numerical study |
| $\vec{m}$ | Magnetization distribution |
| $m$ | Valley field ( $m \equiv \vec{m} \cdot \hat{z}$ ) |
| $m^{(0)}$ | Valley field in the absence of external fields |
| $m^{(1)}$ | External field-induced change in valley field |
| $m_{eff}$ | Electron effective mass |
| $R_{VOI}$ | Valley-orbit interaction strength parameter |
| $S^{(1)}$ | External field-induced source term in the valley field equation |
| $V(x,y)$ | Electrical potential energy |
| $V^{(y)}(y)$ | Electrical potential energy profile in the y direction in a Q1D structure |
| $V^{(\mathrm{control})}(y)$ | Symmetric electrical potential energy profile in the y-direction to control graphene subband structure |
| VMM | Valley magnetic moment corresponding to valley field accumulated over a Q1D structure or a quantum dot |
| $\mathrm{VMM}_{1/2}$ | Valley field accumulated over half width of a Q1D structure |
| $v_F$ | Fermi velocity parameter in the Dirac model |
| $W$ | Nanoribbon width |
| $W_{\mathrm{channel(barrierr)}}$ | Channel (barrier) width in a Q1D structure |
| $W_{\mathrm{x(y)}}$ | Rectangular quantum dot dimension in the x (y) direction |
| $\Delta(x, y)$ | Local bulk gap parameter |
| $\Delta^{(y)}(y)$ | Bulk gap parameter profile in the y direction in a Q1D structure |
| $\Delta^{(0)}$ | The constant term in a space-modulated Δ |
| $\Delta_{channel(barrier)}$ | Bulk gap constant in the channel (barrier) of a Q1D structure |
| $\delta k_x$ | Electric field-induced wave vector difference |
| δΔ | The modulated part in Δ |
| $\mu_B$ | Bohr magneton, used as a unit of magnetic moments in the gapless case |
| $\mu_\tau$ | Valley magnetic moment of the bulk band edge state at valley $\tau$ |
| $\mu_B^*$ | $\lvert \mu_\tau \rvert$, a unit of magnetic moments like $\mu_B$ to use in the gapped case |
| $\mu_{bulk}$ | Valley magnetic moment of a bulk electron state |
| $\mu_{other}$ | Non-valley magnetic moment, e.g., spin magnetic moment |
| $\rho_{A(B)}$ | Electron probability on A (B) site |
| $\rho$ | Electron probability sum, i.e., $\rho_A + \rho_B$ in a unit cell |
| $\rho^{(0)}$ | $\rho$ in the absence of external fields |
| $\rho_{diff}$ | Electron probability difference, i.e., $\rho_A - \rho_B$ in a unit cell |
| $\rho_{diff}{}^{(0)}$ | $\rho_{diff}$ in the absence of external fields |
| $\tau$ | Electron valley index, e.g., +1 / -1 for valley K / K' |
| $\phi_\tau$ | Bulk band edge Bloch state of valley $\tau$ |
| $\psi$ | Total electron wavefunction in the effective-mass description |

# Appendix A

## The variant Schrodinger theory

We illustrate discuss the variant Schrodinger equation in the Q1D case. The discussion can easily be generalized to the quantum dot case.

The structure considered is subject to the modulation of $V^{(y)}_{total}(y)$, where

$$V^{(y)}_{total}(y) = \begin{cases} V^{(y)}(y) & \text{(TMDCs)} \\ V^{(y)}(y) \pm \delta\Delta^{(y)}(y) & \text{(graphene)} \end{cases} \tag{21}$$

with the inclusion of additional contribution $\delta\Delta^{(y)}(y)$ in the graphene case (+ / – for conduction / valence band; $\delta\Delta^{(y)}(y)$ = gap modulation in graphene).

The derivation is based on the effective-mass approximation,

$$\psi(x, y) = \exp(ik_x x) f(y)\phi_\tau(x, y) \tag{22}$$

($f(y)$ = slowly varying envelop function, $\phi_\tau(x, y)$ = band-edge Bloch state at valley $\tau$ ). $f(y)$ satisfies the following effective-mass equation:

$$\frac{1}{2m_{eff}}\left(k_x^2 - \partial_y{}^2\right) f(y) + V^{(y)}_{total} f(y) = (E - E_{\vec{k}=0}) f(y) \tag{23}$$

($E_{\vec{k}=0}$ = corresponding bulk band edge).

Since $f(y)$ is real for a Q1D state, it follows that $\rho(y) = f(y)^2$, and

$$\widehat{H_S}\rho(y) = \left(E - E_{\vec{k}=0}\right)\partial_y \rho(y), \tag{24}$$

$$\widehat{H_S} \equiv \frac{1}{2m_{eff}}\left(k_x^2 - \frac{1}{4}\partial_y{}^2\right)\partial_y + V^{(y)}_{total}(y)\partial_y + \frac{\partial_y V^{(y)}_{total}(y)}{2}$$

Finally, using $m(y) \sim \rho(y)\mu_\tau$ (Eqn. (14)) in Eqn. (24), we obtain the Schrodinger equation for $m(y)$:

$$H_S m(y) = \left(E - E_{\vec{k}=0}\right)\partial_y m(y) \tag{25}$$

To include $B_z(y)$ or the valley-orbit interaction, we make the following replacement in Eqn. (23) as shown in previous studies [28,81]:

$$V^{(y)}_{total}(y) \to V^{(y)}_{total}(y) + V^{(y)}_{field}(y),$$

$$V^{(y)}_{field}(y) = V^{(y)}_{valley-field}(y) + V^{(y)}_{non-valley-field}(y), \tag{26}$$

$$V^{(y)}_{non-valley-field}(y) = -\frac{k_x - A_x(y)/2}{m_{eff}} A_x(y) - \mu_{other} B_z(y), \tag{27}$$

$$V^{(y)}_{valley-field}(y) = \begin{cases} \tau R_{VOI} k_x \partial_y V^{(y)}_{total} - \mu_\tau B_z(y) \\ \text{(TMDCs)} \\ \tau R_{VOI} k_x \left( \partial_y V^{(y)}_{total} - 2\partial_y \Delta \right) - \mu_\tau B_z(y) \\ \text{(graphene)} \end{cases} \quad (28)$$

( $\mu_\tau$ = valley magnetic moment, $\mu_{other}$ = non-valley magnetic moment, $R_{VOI}$ = valley-orbit interaction strength parameter). The same replacement in Eqn. (25) then gives the corresponding Schrodinger equation for $m(y)$ in the case.

# Appendix B

# The variant Klein-Gordon theory in the presence of abrupt asymmetric boundaries

Abrupt asymmetric boundaries in ZGNRs have an important impact on valley fields. For example, as discussed in **Sec. III-2**, it can lead to valley fields with nearly antisymmetric profiles or, equivalently, the suppression of valley magnetic moments near Dirac points. From the theoretical perspective, the foregoing nontrivial manifestation imply a fundamental alteration of valley physics by the boundaries. In particular, as will be shown below, a pseudo vector potential parameter "$A_x^{(BC)}(\tau)$" emerges in the theory as the result of boundary asymmetry.

This Appendix uses zigzag graphene nanoribbons as the example, discusses the corresponding Klein-Gordon valley field equation in the absence of external fields, and illustrates the boundary effect analytically.

**B-1** discusses the emergence of the vector potential. **B-2** derives the Klein-Gordon equation. **B-3** shows the suppression of valley magnetic moment near a Dirac point.

## B-1. Emergence of the pseudo vector potential

The emergence of pseudo vector potential is roughly explained as follows. Basically, the boundary asymmetry across the ZGNR simulates the asymmetry induced by an electric field in the y-direction, and generates in the electron energy a term linear in $k_x$ similar to the valley-orbit-interaction given in Eqn. (18). Such a term implies the existence of a pseudo magnetic field or, equivalently, a vector potential, when compared to the Landau energy term $\frac{k_x}{m_{eff}} A_x$ of an electron in 1D.

For a start, in the absence of external fields, we list below a few useful identities involving the current distribution $j_x$, probability distribution $\rho(y)$, and probability distribution difference $\rho_{diff}(y)$ (E = electron state energy, $\Delta$ = bulk gap parameter, and $\tau$ = valley index):

$$\partial_y j_x = 2\tau\Delta\rho - 2\tau E\rho_{diff} \,, \quad (29)$$

$$j_x = \frac{k_x}{E}\rho + \frac{\tau\partial_y \rho_{diff}}{2E} \,, \quad (30)$$

$$j_x = \frac{k_x}{\Delta}\rho_{diff} + \frac{\tau\partial_y \rho}{2\Delta} \,, \quad (31)$$

all of which can be derived from the Dirac equation.

**Effect of asymmetric boundaries**

We examine the effect of asymmetric, vanishing amplitude boundary condition in the ZGNR, e.g., $F_A(x = W/2) = F_B(x = -W/2) = 0$ in terms of the current distribution $j_x$. Eqn. (30) gives

$$\int j_x dy = \frac{k_x - A_x^{(BC)}(\tau)}{E},$$
$$A_x^{(BC)}(\tau) \equiv -\int \frac{\tau\partial_y \rho_{diff}}{2} dy = -\frac{\tau}{2}\left[\rho_{diff}(W/2) - \rho_{diff}(-W/2)\right] \quad (32)$$

Above, "$k_x / E$" is identical to the free current in Dirac model, and $A_x^{(BC)}(\tau)$ is an asymmetric boundary-induced parameter. As $A_x^{(BC)}(\tau) \propto \left[\rho_{diff}(W/2) - \rho_{diff}(-W/2)\right]$, it serves as a measure of the boundary asymmetry. Moreover, based on the boundary condition of vanishing amplitudes, $\rho_{diff}(W/2) - \rho_{diff}(W/2) = -\left[\rho_B(W/2) + \rho_A(-W/2)\right] \neq 0$ making $A_x^{(BC)}$ nonvanishing.

Eqn. (32) can be used to locate the subband edge, where $\int j_x dy = 0$. It follows that the edge occurs at $k_x = A_x^{(BC)}(\tau)$, implying a valley-dependent wave vector shift of the edge away from the Dirac point ($k_x = 0$). Such a shift suggests the substitution $k_x \to k_x - A_x^{(BC)}(\tau)$ with $A_x^{(BC)}(\tau)$ interpreted as a pseudo vector potential parameter. A variational argument can be applied to obtain the parameter explicitly in terms of the ribbon width $W$. Without going through details of the argument, we provide the expression below:

$$A_x^{(BC)}(\tau) = -\tau \left\langle \rho_A + \rho_B \right\rangle_{average} = -\frac{\tau}{W} \quad (33)$$

(<…>$_{average}$ denotes the spatial average of the expression in bracket) in the case of non-edge states. The result is rigorously consistent with the expectation that the boundary effect on a non-edge state scales with the "surface/volume" ratio, e.g., $1/W$.

**Figure 11** provides a numerical band structure result and shows the second conduction subband in ZGNRs at various ribbon width $W$ 's. It shows that $k_x$ of a subband edge state shifts away by "$\tau / W$" from the Dirac point, in a valley contrasting fashion. The result confirms the presence of $A_x^{(BC)}(\tau)$ as well as Eqn. (33).

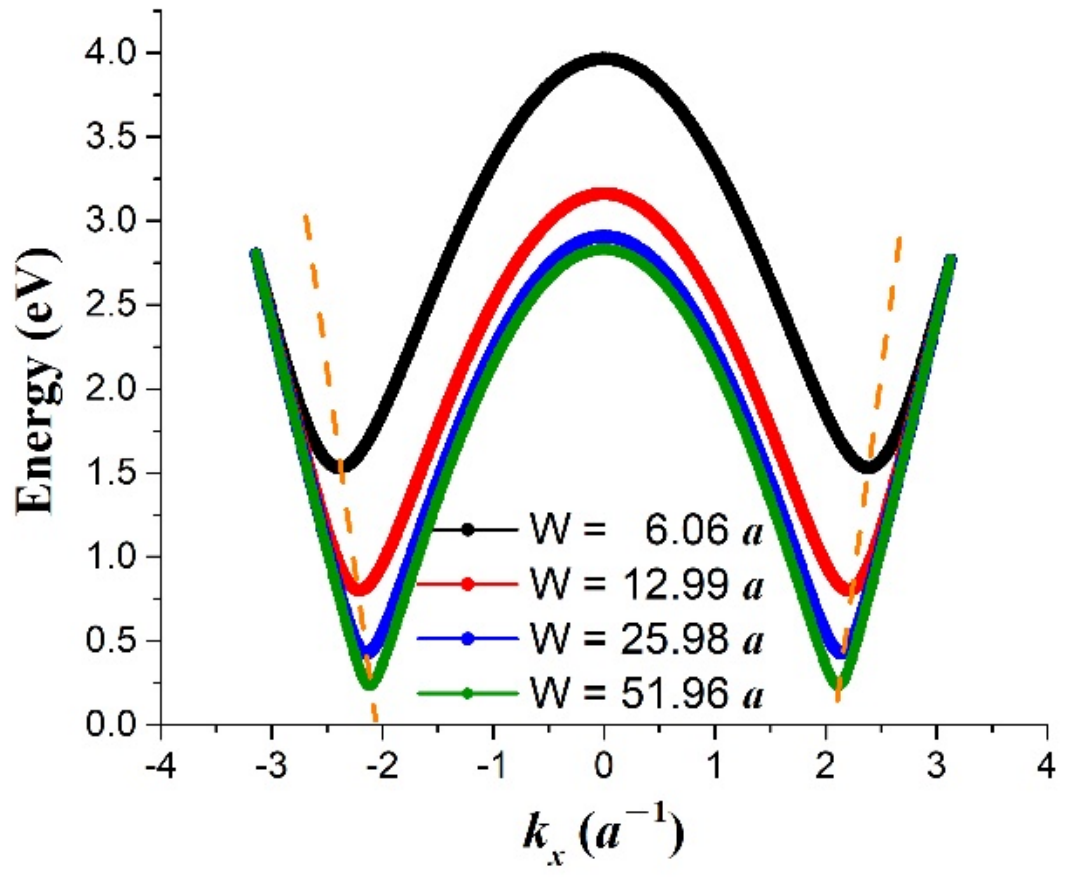

**Figure 11. Abrupt asymmetric boundaries induced band edge shift and corresponding pseudo vector potential** The second conduction band at various ribbon width $W$'s. The band edge moves away from Dirac point when $W$ decreases. The ZGNR is characterized by bulk gap parameter $\Delta = 0.1\ eV$ .

## B-2. The Klein-Gordon equation

Apart from shifting the subband edge, the pseudo vector potential $A_x^{(BC)}(\tau)$ has a nontrivial effect on the valley field $m(y)$ as shown below. Combining Eqns. (7) and (32) gives the following magnetization current distribution

$$\partial_y m(y) = \frac{\tau\partial_y \rho_{diff}}{2E} + \frac{A_x^{(BC)}}{E}\rho \ , \qquad (34)$$

where the second term on the right hand side explicitly shows the induction of a magnetization current by $A_x^{(BC)}(\tau)$ . Such an effect needs to be accounted for when deriving the valley field equation.

### The equation

Eqns. (29)-(31) and (34) constitute a set of coupled equations for the four field variables $m(y)$ , $j_x(y)$, $\rho(y)$, and $\rho_{diff}(y)$ . Elimination of variables leads to the following equation for $m(y)$ :

$$-\frac{1}{4}\partial_y^{\,2}\left[\partial_y^{\,2}m(y)\right] + \left(k_x^2+\Delta^2\right)\left[\partial_y^{\,2}m(y)\right] = E^2\left[\partial_y^{\,2}m(y)\right]. \qquad (35)$$

### Boundary conditions

The foregoing valley field equation is fourth order and so requires four boundary conditions. These conditions follow from the requirement of vanishing $m(y)$ and $j_x(y)$ at $y = \pm W/2$. Again, Eqns. (29)-(31) and (34) can be applied to convert these conditions into the following ones in terms of $m(y)$ and its derivative:

$$m(\pm W/2) = 0 \qquad (36)$$

$$\begin{aligned}&\left[\pm k_x(k_x - A_x^{(BC)}) - \mathrm{E}(\Delta \pm \mathrm{E})\right]\partial_y m(\pm W/2)\\ &-\frac{1}{2}(k_x - A_x^{(BC)})\partial_y^2 m(\pm W/2) = 0.\end{aligned} \qquad (37)$$

Eqns. (35)-(37) constitute the Klein-Gordon theory of valley fields in ZGNRs.

## B-3. Parity of valley fields

Last, $m(y)$ of a subband edge state, where $k_x = A_x^{(BC)}$, is shown to have an odd parity.

**Proof:**

i) When $k_x = A_x^{(BC)}$, Eqn. (37) reduces to $\partial_y m\left(\pm\frac{W}{2}\right) = 0$. Therefore, the valley field equation (35) and boundary conditions (36) and (37) are all mirror symmetric with respect to the reflection y ↔ -y. So, the solution $m(y)$ has parity symmetry.

ii) Eqns. (29) – (31) and (34) can be used to show the following inequality

$$\begin{aligned}&\partial_y^{\,2}m(W/2) - \partial_y^{\,2}m(-W/2)\\ &= \frac{4}{E}(\Delta^2 - E^2)A_x^{(BC)} \ , \\ &\neq 0\end{aligned} \qquad (38)$$

which excludes the alternative of even parity in $m(y)$. In conclusion, $m(y)$ is odd.

**Q.E.D.**

As valley magnetic moments are sum totals of valley fields in space, the foregoing result equivalently states vanishing valley magnetic moments at and largely suppressed valley magnetic moments near Dirac points, in ZGNRs.

# Appendix C

# The variant Klein-Gordon theory

# in the presence of

# space-dependent fields

This Appendix illustrates the mathematical development of valley field mechanics in space-dependent electric and magnetic fields, in the Dirac two-band model using Q1D graphene structures confined with barriers as the example. The field-free, Klein-Gordon theory is presented firstly, followed by an inclusion of external fields in the theory and the discussion of local valley-external field interactions, in the linear response regime.

Important connections exist among the valley field $m(y)$ , current distribution $j_x(y)$, probability distribution $\rho(y)$, and probability distribution difference $\rho_{diff}(y)$ , as expressed in Eqns. (7) and (8). The Appendix follows the strategy of taking $\rho_{diff}(y)$ as the auxiliary field, building the corresponding auxiliary field equation, and based on it, developing the valley field equation.

**C-1** formulates the auxiliary field equation in the presence of external fields. **C-2** presents the field-free, Klein-Gordon valley field theory. **C-3** discusses the valley field theory in the presence of external fields as well as local valley -external field interactions, in the linear response regime.

## C-1. The auxiliary field equation

For a start, we list below important identities including Eqns. (7) and (8), which involve $m(y)$, $j_x(y)$, $\rho_{diff}(y)$, and $\rho(y)$ and can be derived from the Dirac equation in the presence of external fields (E = electron state energy, $V^{(y)}$ = electrical potential energy, $A_x^{(y)}$ = vector potential, $\Delta^{(y)}$ = modulated gap, $\tau$ = valley index):

$$\begin{aligned}
&\partial_y j_x(y) = 2\tau\Delta^{(y)}(y)\rho(y) - 2\tau\left[E - V^{(y)}(y)\right]\rho_{diff}(y),\\
&j_x(y) = \left(\frac{k_x - A_x(y)}{E - V^{(y)}(y)}\right)\rho(y) + \left(\frac{\tau}{2\left[E - V^{(y)}(y)\right]}\right)\partial_y \rho_{diff}(y),\\
&j_x(y) = \left(\frac{k_x - A_x(y)}{\Delta^{(y)}(y)}\right)\rho_{diff}(y) + \left(\frac{\tau}{2\Delta^{(y)}(y)}\right)\partial_y \rho(y),\\
&\partial_y m(y) = j_x(y) - \left[\int \langle j_x \rangle dy\right]\rho(y).
\end{aligned} \tag{39}$$

Above identities constitute a set of simultaneous equations for $m(y)$, $j_x(y)$, $\rho_{diff}(y)$, and $\rho(y)$. Elimination of variables is applied, giving

$$0 = \left[-h_0 \partial_y \circ \hat{h}_2 + (\partial_y h_0)\hat{h}_2 + h_0 \hat{h}_1\right]\rho_{diff}(y) \ , \tag{40}$$

where $h_0$, $h_1$, and $h_2$ are operators defined below:

$$\begin{aligned}
&h_0(V^{(y)}, A_x, E)\\
&\equiv 1 - \frac{\tau}{2\Delta^{(y)}} d_y\left[\frac{(k_x - A_x)}{E - V^{(y)}}\right] - \left[\frac{(k_x - A_x)}{E - V^{(y)}}\right]^2,
\end{aligned}$$

$$\begin{aligned}
&\hat{h}_1(V^{(y)}, A_x, E)\rho_{diff}\\
&\equiv d_y\left[\frac{(k_x - A_x)}{E - V^{(y)}}\right]\frac{(k_x - A_x)}{\Delta^{(y)}}\rho_{diff}\\
&+\left[2\tau\Delta^{(y)} - \frac{d_y(k_x - A_x)}{E - V^{(y)}}\right]\frac{\tau}{2(E - V^{(y)})} d_y \rho_{diff}\\
&+\frac{\tau(k_x - A_x)}{2(E - V^{(y)})^2} d_y^{\,2} \rho_{diff}
\end{aligned} ,$$

$$\begin{aligned}
&\hat{h}_2(V^{(y)}, A_x, E)\rho_{diff}\\
&\equiv \left[\frac{E - V^{(y)}}{\Delta^{(y)}} - \frac{(k_x - A_x)^2}{\Delta^{(y)}(E - V^{(y)})}\right]\rho_{diff}\\
&+\left[\frac{1}{4\Delta^{(y)}} d_y\left(\frac{1}{E - V^{(y)}}\right) + \frac{\tau(k_x - A_x)}{2(E - V^{(y)})^2}\right] d_y \rho_{diff}\\
&+\frac{1}{4\Delta^{(y)}(E - V^{(y)})} d_y^{\,2} \rho_{diff}
\end{aligned} . \tag{41}$$

For graphene Q1D structures confined with barriers, we have the following boundary condition:

$$\rho_{diff}(y \to \pm\infty) \to 0. \tag{42}$$

Eqns. (40) and (42) constitute the auxiliary field equation in the presence of external fields.

## C-2. Field-free Klein-Gordon theory

In the absence of external fields, we write

$$\begin{aligned}
&\rho(y) = \rho^{(0)}(y),\\
&\rho_{diff}(y) = \rho_{diff}^{(0)}(y),\\
&j_x(y) = j_x^{(0)}(y),\\
&m(y) = m^{(0)}(y),\\
&E = E^{(0)}.
\end{aligned} \tag{43}$$

Eqn. (41) can be applied to express $\rho_{diff}^{(0)}$, $\rho^{(0)}$, and $j_x^{(0)}(y)$ in terms of $E^{(0)}$ and $m^{(0)}(y)$:

$$\begin{aligned}
&\rho_{diff}^{(0)}\left[m^{(0)}\right] = 2\tau E^{(0)} m^{(0)}(y),\\
&\rho^{(0)}\left[m^{(0)}\right] = 2\tau E^{(0)}\left[h_0^{-1}\hat{h}_2\right]^{(0)} m^{(0)}(y)\\
&= \frac{2\tau E^{(0)2}}{\Delta^{(y)}(y)} m^{(0)}(y) + \frac{k_x}{E^{(0)}}\partial_y m^{(0)}(y)\\
&+ \frac{\tau}{2\Delta^{(y)}(y)}\partial_y^{\,2} m^{(0)}(y),\\
&j_x^{(0)}\left[m^{(0)}\right] = \frac{k_x}{E^{(0)}}\rho^{(0)}\left[m^{(0)}\right] + \partial_y m^{(0)}(y)
\end{aligned} \tag{44}$$

$\left([...]^{(0)} \equiv [...]\big|_{V^{(y)}=0, A_x=0, E=E^{(0)}}\right)$.

Expressions provided above are useful in deriving the valley field equation below and in the discussion of field effects in **C-3**.

Combining the expression of $\rho_{diff}^{(0)}\left[m^{(0)}\right]$ provided above and Eqns. (40) and (42) leads to the following valley field equation:

$$H_{KL} m^{(0)}(y) = E_0^{\,2} m^{(0)}(y) \ ,$$

$$H_{KL}m^{(0)}(y)$$
$$\equiv\left(k_x^2-\frac{1}{4}\partial_y^2\right)m^{(0)}(y)+\Delta^{(y)}(y)\int_{-\infty}^{y}\Delta^{(y)}(y)\left(\partial_y m^{(0)}(y)\right)dy,$$
$$m^{(0)}(y\to\pm\infty)\to 0 \quad , \tag{45}$$

with $H_{KL}$ the field-free Klein-Gordon type operator.

### C-3. Effects of external fields

In the presence of $V^{(y)}$ or $A_x$, we derive the field effects in the linear response regime as follows. We linearize all relevant variables, e.g.,

$$\rho(y)\approx\rho^{(0)}(y)+\rho^{(1)}(y),$$
$$\rho_{diff}(y)\approx\rho_{diff}^{(0)}(y)+\rho_{diff}^{(1)}(y),$$
$$j_x(y)\approx j_x^{(0)}(y)+j_x^{(1)}(y),$$
$$E\approx E^{(0)}+E^{(1)},$$
$$m(y)\approx m^{(0)}(y)+m^{(1)}(y), \tag{46}$$

where $\rho^{(1)}$, $\rho_{diff}^{(1)}$, $j_x^{(1)}$, $E^{(1)}$, and $m^{(1)}$ are field-induced responses linear in the fields. In particular, $E^{(1)}$ is given by the following first-order perturbation-theoretic expression in terms of the field-free solution:

$$E^{(1)}=\left\langle V^{(y)}\right\rangle_0-\langle A_x\rangle_0,$$
$$\left\langle V^{(y)}\right\rangle_0\equiv\int_{-\infty}^{\infty}V^{(y)}(y)\rho^{(0)}(y)dy, \tag{47}$$
$$\langle A_x\rangle_0\equiv\int_{-\infty}^{\infty}A_x(y)j_x^{(0)}(y)dy$$

($\langle ...\rangle_0$ denotes expectation value with respect to the field-free solution).

**Local valley-orbit and valley-Zeeman interactions**

Substituting the expressions provided in Eqn. (44) into Eqn. (47) and collecting the contributions to $\left\langle V^{(y)}\right\rangle_0$ and $\langle A_x\rangle_0$ from the valley-dependent terms, we obtain the following leading-order local valley-external field interaction energy

$$E_{valley-field}$$
$$\approx\int_{-\infty}^{\infty}\frac{k_x}{E^{(0)}}\partial_y V^{(y)}m^{(0)}(y)dy-\int_{-\infty}^{\infty}B_z(y)m^{(0)}(y)dy \tag{48}$$

giving

$$\frac{k_x}{E^{(0)}}\partial_y V^{(y)}m^{(0)}(y) \tag{49}$$

as the *local valley-orbit interaction* and

$$-B_z(y)m^{(0)}(y) \tag{50}$$

as the *local valley-Zeeman interaction.*

**Theory in the linear response regime**

In the linear response regime, the field equation for $m^{(1)}(y)$ can be obtained by linearizing Eqns. (39) and (40), which yields the following Klein-Gordon equation with a source term:

$$\left[H_{KL}-E^2\right]m^{(1)}(y)=s^{(1)}(y;V^{(y)},A_x). \tag{51}$$

This equation can be solved to provide $m^{(1)}(y)$ in terms of the external fields. Above, the source term $s^{(1)}(y;V^{(y)},A_x)$ is an expression linear in $V^{(y)}$ and $A_x$, given below in terms of the field-free solution:

$$s^{(1)}(y;V^{(y)},A_x)$$
$$\equiv\frac{\tau}{2}\Delta^{(y)}(y)\int_{-\infty}^{y}\left[-\partial_y\circ\hat{h}_2+\hat{h}_1\right]^{(0)}d^{(1)}(y;V^{(y)},A_x)\,dy,$$
$$-\frac{\tau}{2}\Delta^{(y)}(y)\int_{-\infty}^{y}\frac{1}{\left[h_0\right]^{(0)}}\left[-h_0\partial_y\circ\hat{h}_2+(\partial_y h_0)\hat{h}_2+h_0\hat{h}_1\right]^{(1)}m^{(0)}(y)\,dy$$

$$d^{(1)}(y;V^{(y)},A_x)$$
$$\equiv\frac{1}{E_0}\int_{-\infty}^{y}\left[n_V^{(1)}(y)+n_{A_x}^{(1)}(y)\right]dy,$$

$$n_V^{(1)}(y)$$
$$\equiv V^{(y)}(y)\partial_y m_0(y)$$
$$+\left\{\frac{k_x\left[V^{(y)}(y)-\left\langle V^{(y)}\right\rangle_0\right]}{E_0}-\int_{-\infty}^{\infty}(\partial_y m_0)V^{(y)}(y)dy\right\}\rho_0(y),$$

$$n_{A_x}^{(1)}(y)\equiv\left[\langle A_x\rangle_{0'}-A_x(y)\right]\rho_0(y) \tag{52}$$

($\langle A_x\rangle_{0'}\equiv\int_{-\infty}^{\infty}A_x(y)\rho_0(y)dy$ ; $[\ldots]^{(0)}$ and $[\ldots]^{(1)}$ denote the field-independent and linear-in-field parts of the expression in bracket, respectively).

## Appendix D

## Effects of valley mixing

When intervalley scattering exists, it mixes the opposite valley states and, thus, reduces the valley field amplitude. Such scattering is present in graphene structures with armchair boundaries. In TMDC structures, because of spin-valley and spin-valley-orbital coupling, armchair edge scattering alone is not sufficient to couple valleys, unless a spin-flipping mechanism is simultaneously induced, for example, by a vertical electric field through the spin-orbit interaction. [81]

When intervalley coupling is present, the two valleys are evenly mixed giving a vanishing cell-orbital magnetic moment. A vertical magnetic field $B_z$ can be introduced to break the even mixing and polarize the state via the valley-Zeeman interaction, with the polarization dependent on the competition between the intervalley coupling and the valley-Zeeman interaction.

**D-1** presents numerical results of valley fields in the presence of valley mixing. **D-2** describes an analytical theory of valley fields that accounts for the effect of valley mixing.

## D-1. Numerical results

**Figure 12** presents valley fields of top two valence subbands in the armchair graphene nanoribbon with ribbon width W = 19 *a*. The polarizing magnetic field is given by $\mu_B^* B_z$ = -235 meV. **(a)** shows the two subbands. **(b)** shows valley magnetic moment (VMM) vs. $k_x$ for each subband.

**Armchair graphene nanoribbon**

(a)

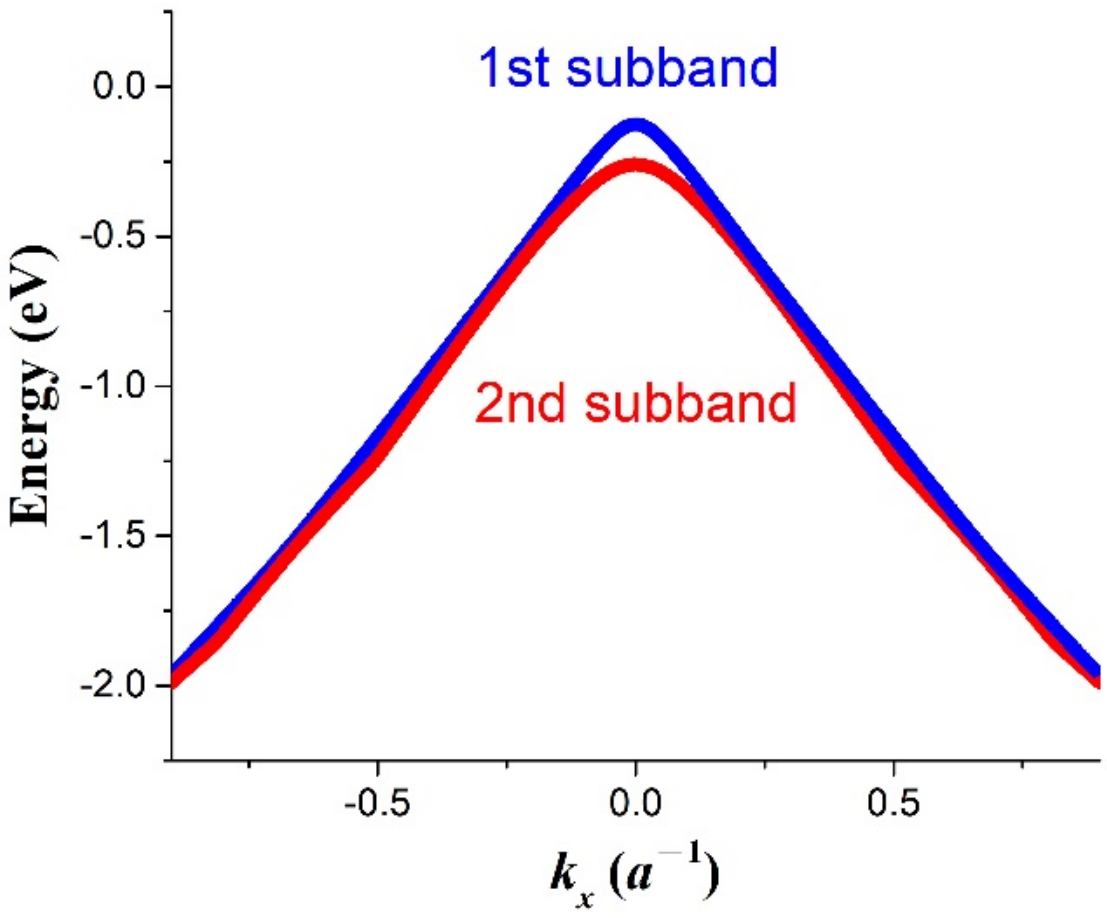

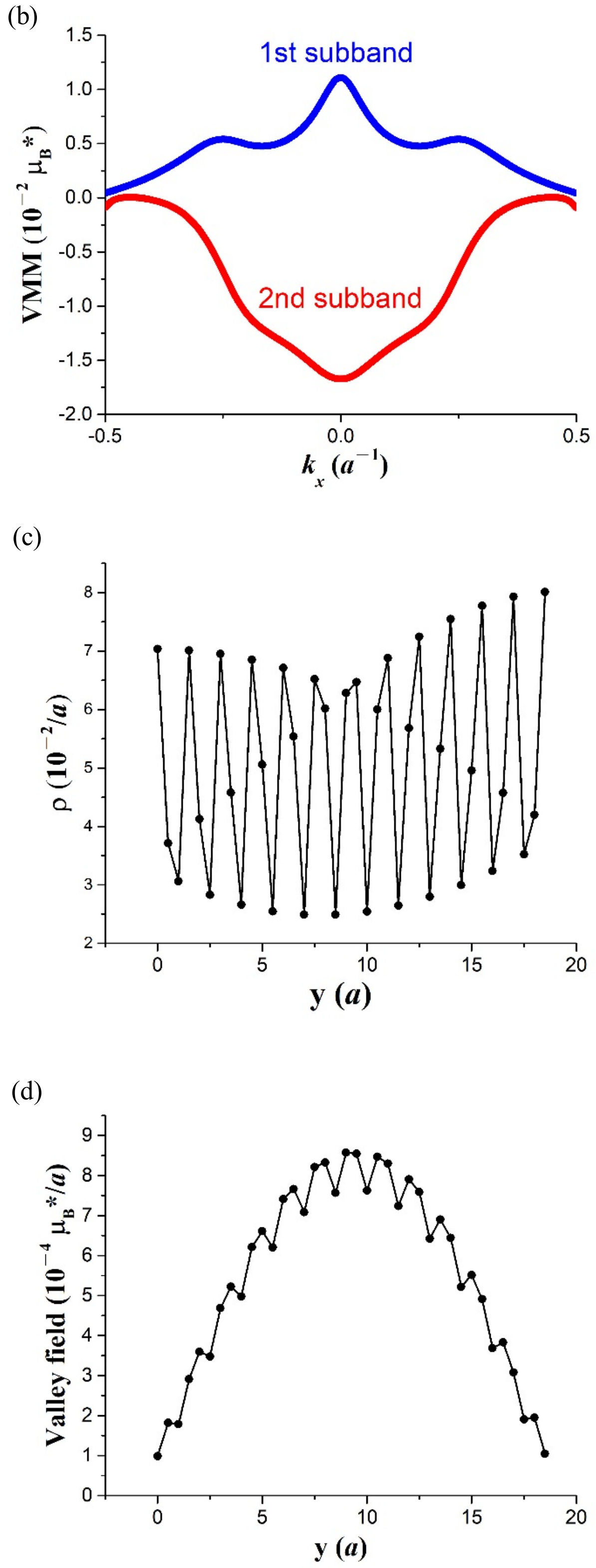

**Figure 12. Magnetic field polarized valley field in AGNR** The ribbon has ribbon width W = 19 *a* and bulk gap parameter Δ = 0.02 eV. **(a)** Top and second valence subbands. **(b)** Valley magnetic moment (VMM) vs $k_x$ of the bands in **(a)**. **(c)** and **(d)** are raw data of $\rho(y)$ and valley field, respectively, without coarse grain averaging, of the top subband state at $k_x$ = -0.018 $a^{-1}$, both of which show rapid oscillations – the signature of valley mixing.

**(c)** and **(d)** present raw data of probability distribution ( $\rho(y)$ )and valley field, respectively, without coarse grain averaging, of the top subband state at $k_x$ = -0.018 $a^{-1}$.

With the abrupt armchair edges, a strong valley mixing is induced, which has several implications: i) it limits the valley field and valley magnetic moment as well, resulting in $|\mu_{AGNR}|$< 0.02 $\mu_B^*$ in the two subbands as shown in **(b)**; ii) it breaks the probability-valley field correlation, as can be verified by comparing **(c)** and **(d)**; and iii) it leads to the inter-valley interference − rapid oscillations manifested in both $\rho(y)$ and valley field shown in **(c)** and **(d)**

**Figure 13** features valley-unpolarized electron states at $k_x$ = - 0.011 $a^{-1}$ in two armchair $WSe_2$ Q1D structures, one of which has ribbon width W = 25 *a* and is confined by abrupt edges (in **(a)**), and the other is confined by barriers with channel width $W_{channel}$ = 25 *a* and barrier width $W_{barrier}$ = 35 *a* (in **(b)**). Both structures are subject to a vertical electric field $E_z$ = 10 mV/*a* to effect intervalley scattering. Both valley fields and probability distributions are presented. As shown in the figure, the scattering results, in the two cases, quantitatively similar, antisymmetric valley fields with small amplitudes and corresponding vanishing valley magnetic moments, as well as breaking of probability-valley field correlation. As for the antisymmetric feature shown in the valley field profile, the small amplitude implies its being some secondary effect in contrast to that observed in the ZGNR case in **Figures 7 and 8**. A brief note of mechanisms for such secondary effect is given in **D-2** when discussing the analytical theory.

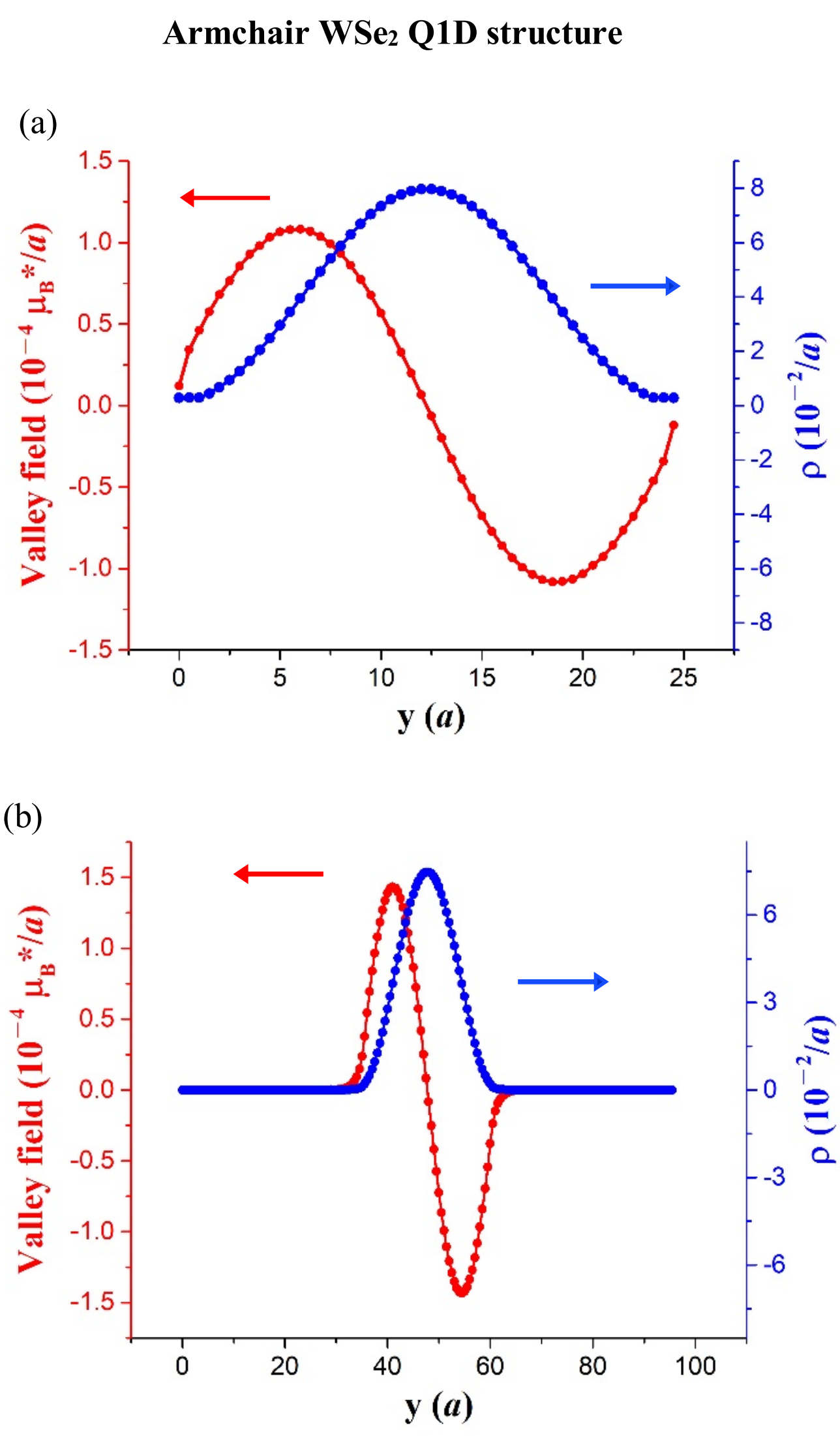

**Figure 13. Unpolarized valley fields in armchair $WSe_2$ Q1D structures** Valley fields and $\rho(y)$ 's of unpolarized top valence subband states at $k_x$ = - 0.011 $a^{-1}$, in two armchair $WSe_2$ Q1D structures. The structure considered in **(a)** is confined by abrupt edges with ribbon width W = 25 *a*. The structure considered in **(b)** is confined by barriers with channel width $W_{channel}$ = 25 *a* and barrier width $W_{barrier}$ = 35 *a*. Both structures are subject to a vertical electric field $E_z$ = 10 mV/*a*. In the case of barrier confinement, a negative on-site energy (-1 eV) is applied to form the barrier.

## D-2. Analytical theory

Below, we consider the weak intervalley coupling limit where

$$E_{intervalley\ coupling} \ll E_{quantization} \tag{53}$$

( $E_{intervalley\ coupling}$ = intervalley coupling, $E_{quantization}$ = quantization energy) in Q1D TMDC and graphene structures confined with barriers, and present an analytical treatment of valley mixing in the regime based on a perturbation theory. Away from the regime, valley mixing starts to suppress *m* ending up with an insignificant residual value. Such cases occur in QDs and AGNRs with abrupt boundaries and are best studied by a numerical approach as illustrated in **D-1**.

### Valley-degenerate case

$B_z(y)$= 0. Ignore the intervalley coupling first and focus on the zeroth-order solution - $m_{K(K')}^{(0)}(y)$ for K(K') valley field. The fields are governed by the Schrodinger Eqn. (25) or Klein-Gordon Eqn. (45), with corresponding energy $E_{K(K')}^{(0)}$. Due to the time reversal symmetry, $E_K^{(0)} = E_{K'}^{(0)}$ and $m_{K'}^{(0)}(y) = -m_K^{(0)}(y)$ . Intervalley coupling mixes the two fields evenly, giving the sum field

$$m(y) = 0. \tag{54}$$

The result is valid when higher-order effects, e.g., valley-orbit interaction or trigonal band warping around a Dirac point are neglected. When they are included, a small yet finite difference exists between $|m_K^{(0)}(y)|$ and $|m_{K'}^{(0)}(y)|$ giving a residual *m*(*y*), as shown in **Figure 13** with a full tight-binding calculation.

### Valley-polarized case

A vertical magnetic field $B_z(y)$ is applied to lift the degeneracy and polarize the state. Consider $B_z(y)$ in the linear regime. Described below is a one-parameter ( $E_{intervalley\ coupling}$ ) , two-state perturbation theory of the linear response $m^{(1)}(y)$ .

In the Hilbert space of the two field-free, degenerate solutions, the two-state Hamiltonian is given by

$$H_{armchair} \approx \begin{pmatrix} E_{valley-Zeeman} & E_{intervalley\ coupling} \\ E_{intervalley\ coupling} & -E_{valley-Zeeman} \end{pmatrix} \qquad (55)$$

where the valley-Zeeman $E_{valley-Zeeman} = -\int_{-\infty}^{\infty} B_z(y) m_K^{(0)}(y) dy$ .

$m^{(1)}(y)$ is determined by the competition between valley-Zeeman interaction and intervalley coupling described in Eqn. (55). Then

$$m^{(1)}(y) \approx (|\alpha_K|^2 - |\beta_{K'}|^2) m_K^{(0)}(y) \qquad (56)$$

Above, $(\alpha_K, \beta_{K'})^t$ denotes the eigenstate of $H_{armchair}$ . The valley field has been taken to be approximately valley diagonal, as the off-diagonal part varies rapidly as $\exp(\pm\ i\ 2K\ y)$ and vanishes in a coarse grain average.